\documentclass[prc,article]{revtex4-1}
\usepackage{graphicx}
\usepackage{amsmath,amssymb}
\usepackage{hyperref}
\usepackage{color}
\usepackage{xcolor}

\colorlet{darkgreen}{green!50!black}
\colorlet{darkblue}{blue!60!black}
\colorlet{darkred}{red!80!black}

\begin{document}

\title{ {\it Ab initio} no-core solutions for $^6$Li }

\author{Ik Jae Shin}
\email{geniean@ibs.re.kr}
\affiliation{Rare Isotope Science Project, Institute for Basic Science, Daejeon 34037, Korea}

\author{Youngman Kim}
\email{ykim@ibs.re.kr}
\affiliation{Rare Isotope Science Project, Institute for Basic Science, Daejeon 34037, Korea}

\author{Pieter Maris}
\email{pmaris@iastate.edu}
\affiliation{Department of Physics and Astronomy, Iowa State University, Ames, IA 50011, USA}

\author{James P. Vary}
\email{jvary@iastate.edu}
\affiliation{Department of Physics and Astronomy, Iowa State University, Ames, IA 50011, USA}

\author{Christian Forss\'en}
\email{christian.forssen@chalmers.se}
\affiliation{Department of Physics, Chalmers University of Technology, SE-412 96 G\"oteborg, Sweden}

\author{Jimmy Rotureau}
\email{rotureau@nscl.msu.edu}
\affiliation{Department of Physics, Chalmers University of Technology, SE-412 96 G\"oteborg, Sweden}
\affiliation{National Superconducting Cyclotron Laboratory, Michigan State University, East Lansing, MI 48824, USA}

\author{Nicolas Michel}
\email{Nicolas.Michel@ganil.fr}
\affiliation{Grand Acc\'el\'erateur National d'Ions Lourds (GANIL), 
CEA/DSM - CNRS/IN2P3, BP 55027, F-14076 Caen Cedex, France}
\affiliation{National Superconducting Cyclotron Laboratory, Michigan State University, East Lansing, MI 48824, USA}

\date{\today}

\begin{abstract}
We solve for properties of $^6$Li in the {\it ab initio} No-Core Full Configuration approach and we separately solve
for its ground state and  $J^{\pi}=2_{2}^{+}$ resonance  with the Gamow Shell Model in the Berggren basis. 
We employ both the JISP16 and chiral NNLO$_{\rm opt}$ realistic nucleon-nucleon interactions and 
investigate the ground state energy, excitation energies, point proton root-mean-square radius
and a suite of electroweak observables.
We also extend and test methods to extrapolate the ground state
energy, point proton root-mean-square radius, and electric
quadrupole moment.
We attain improved estimates of these observables in the No-Core Full Configuration approach 
by using basis spaces up through $N_{\rm max}=18$ that enable more definitive comparisons with experiment.
Using the Density Matrix Renormalization Group approach with the JISP16 interaction, we find that we can significantly improve the convergence of the Gamow Shell Model treatment of the $^6$Li
ground state  and $J^{\pi}=2_{2}^{+}$ resonance by adopting a natural orbital single-particle basis.

\end{abstract}

\maketitle

\flushbottom

\section{Introduction}\label{aba:sec1}
A frontier problem in theoretical physics is to understand quantitatively how nuclei, including their structure
and reactions, arise from interacting nucleons based on underlying theories such as quantum chromodynamics (QCD).
Since the interaction among the nucleons inside a nucleus is dominated by the strong interaction, which is non-perturbative
in the low-energy regime relevant for nuclear physics, conventional approaches
based on perturbation theory are not applicable. Therefore, in order to preserve predictive power, it is essential to develop analytic and numerical non-perturbative approaches to solve for nuclear properties using fundamental interactions based on QCD.
Several promising {\it ab initio} methods have been developed for nuclear structure and reactions based on nucleon-nucleon (NN), three-nucleon (3N) and possibly multi-nucleon interactions obtained from either meson exchange theory or an effective field theory of QCD.
At the same time, realistic NN interactions tuned to fit properties of few-nucleon systems serve as valuable benchmarks.
Advances  in high-performance computing have also
played a major role in the upsurge in the non-perturbative many-body theory with these
realistic inter-nucleon interactions.

In this work, we investigate the properties of $^6$Li in the framework of  the {\it ab initio} No-Core Full
Configuration (NCFC) approach~\cite{Maris:2008ax} with the JISP16~\cite{Shirokov:2003kk,Shirokov:2004ff,Shirokov:2005bk} and chiral NNLO$_{\rm opt}$~\cite{Ekstrom:2013kea} potentials
approaching convergence in the largest basis spaces reported to date.
This work is the first to
report results for excited states of $^6$Li based on the chiral  NNLO$_{\rm opt}$ which is derived within a low-energy
effective field theory based on QCD.
We consider both natural and unnatural parity states of $^6$Li. For an earlier study of three lithium isotopes
including $^6$Li within the  {\it ab initio} NCFC method with  the JISP16 potential in smaller basis spaces,
we refer to the work by Cockrell {\it et al.}~\cite{Cockrell:2012vd}.
The NCFC approach is a variation of the  {\it ab initio} No-Core Shell Model (NCSM) with a few distinctive features:
the use of bare interactions, extrapolation to the infinite matrix limit, and uncertainty estimation through the extrapolation.
For overviews of the NCFC approach including JISP16 results for light nuclei, see Refs.~\cite{Maris:2013poa,Shirokov:2014rvw}.
For the NCSM and its applications in light nuclei, including $^6$Li with JISP16 and other interactions,
we refer to a recent review~\cite{Barrett:2013nh}.
We also investigate the ground state and $J^{\pi}=2_{2}^{+}$ resonance of $^6$Li within the Gamow Shell Model (GSM)~\cite{gsm_rev}
with a many-body basis constructed from a single-particle (s.p.) Berggren 
ensemble~\cite{Berggren:1968zz}. Using the  iterative Density Matrix Renormalization Group (DMRG)~\cite{dmrg} method to
solve the GSM Hamiltonian, we show how the convergence rates of both states are significantly improved
when we introduce natural orbitals for the s.p.~states.

\section{Ab initio no core full configuration approach}\label{NCFC}
We briefly introduce the  {\it ab initio} NCFC method.
The Hamiltonian of $A$ nucleons is given by
\begin{equation}
H_A= \frac{1}{A} \sum_{i<j}\frac{({\bf p}_i-{\bf p}_j )^2}{2m} + \sum_{i<j} V_{\textrm{NN},ij}+\sum_{i<j} V_{\textrm{C},ij}\, ,\label{nH}
\end{equation}
where $m$ is the nucleon mass, $V_\textrm{NN}$ ($V_\textrm{C}$)
is the two nucleon (Coulomb) interaction and sums run over all pairs of nucleons.
We also include the harmonic oscillator (HO)
center of mass (CM) Hamiltonian with a Lagrange multiplier, following Refs.~\cite{Maris:2008ax,Cockrell:2012vd},
to eliminate spurious CM excited states and insure factorization
of low-lying eigenvectors into an internal state vector times a CM state vector.  This CM state vector is in the lowest
eigenstate of the HO for CM motion.

For the NN interactions, we adopt the JISP16  
and chiral NNLO$_{\rm opt}$ potentials 
since both describe the NN scattering data and deuteron properties with high accuracy. In addition both were developed
with some effort to reduce the contributions of many-body interactions in light nuclei.

The JISP16 potential is based on the J-matrix inverse scattering approach so that it provides
an excellent description of the phase shifts.  It yields $\chi^2\approx 1$ per degree of freedom
for the $np$ phase shifts at laboratory energies up to $350$~MeV.
JISP16 was developed using phase-shift equivalent
unitary transformations to fit selected experimental data for light nuclei up to $A=16$~\cite{Shirokov:2005bk}.

Chiral effective field theory is a promising theoretical
approach to obtain a quantitative description of the nuclear force from first
principles~\cite{Machleidt:2011zz}.
We adopt the chiral NNLO$_{\rm opt}$ which is obtained by
optimizing the chiral Next-to-Next-to-Leading Order
(NNLO) potential with POUNDERS (Practical Optimization Using No
Derivatives for Squares)~\cite{Ekstrom:2013kea}.
This chiral NNLO$_{\rm opt}$ yields $\chi^2\approx 1$ per degree
of freedom with respect to NN scattering data for laboratory
energies roughly up to $125$~MeV. Furthermore, the
effects of 3N interactions on the properties
of $A=3,4$ nuclei are smaller than previously available parameterizations of
chiral nuclear forces.

With the nuclear Hamiltonian specified above, we solve the $A$-body Schr\"odinger equation
\begin{equation}
H_A \Psi ({\bf r}_1,{\bf r}_2, \dots, {\bf r}_A)=E \Psi ({\bf r}_1,{\bf r}_2, \dots, {\bf r}_A)\, ,
\end{equation}
where  the $A$-body wave function is given by a linear combination of Slater determinants $\Phi_i$
\begin{equation}
 \Psi ({\bf r}_1,{\bf r}_2, \dots, {\bf r}_A)=\sum_{i} c_i \Phi_i ({\bf r}_1,{\bf r}_2, \dots, {\bf r}_A)\, .
\end{equation}
The Slater determinant $\Phi_i$ is the antisymmetrized product of s.p.~wave functions $\phi_a ({\bf r})$, where
$a$ stands for the quantum numbers of a s.p.~state. We use the HO basis for the s.p.~wave
function in our NCFC calculations. In general, to obtain the ground state energy 
and other observables as close as possible
to the exact results we work in the largest feasible basis spaces.  Of course, an infinite basis
space is required for the exact results.
For practical calculations, we truncate the basis space and investigate the trends towards convergence
as the truncation is systematically relaxed.
In the NCFC framework, we adopt the $N_{\rm max}$ truncation scheme, where we retain all HO configurations
where the total number of quanta above the minimum HO energy configuration for the given nucleus is defined by $N_{\rm max}$.  Even values of $N_{\rm max}$ correspond to states with the same parity as the minimum energy HO
configuration (the $N_{\rm max}=0$ configuration) and are called the ``natural'' parity states.  Odd values of
$N_{\rm max}$ then correspond to states with ``unnatural'' parity.
These calculations are performed with the code
MFDn~\cite{Sternberg:2008:ACI:1413370.1413386,DBLP:journals/procedia/MarisSVNY10,CPE:CPE3129},
a hybrid MPI/OpenMP configuration interaction code for  {\it ab initio} nuclear structure
calculations.

The paper is organized as follows.  We first present in Sec.~III our extrapolation methods that are based on the calculated dependence on basis space parameters.  We then present a
synopsis of the application of these methods to the ground state
energy, point proton root-mean-square (rms) radius, and electric quadrupole moment.
Sec.~IV is devoted to presentations of the detailed NCFC results for a range of $^6$Li properties using the chosen
Hamiltonians. 
In Sec.~V we show our Gamow Shell Model results with the DMRG approach for the $^6$Li ground state and  $J^{\pi}=2_{2}^{+}$ resonance for a variety
of basis space selections in order to demonstrate the superior results obtained in the natural orbital basis.
We present our summary in Sec.~VI.

\section{Extrapolation methods}

Our procedure for each observable is to solve for that observable as a
function of $N_{\rm max}$ and as a function of the HO basis parameter
$\hbar\Omega$.  We then examine the sensitivity of that observable to
these basis space parameters ($N_{\rm max}$, $\hbar\Omega$) in order
to gauge convergence.  For selected observables, we employ the
systematic dependence on these basis space parameters to develop and
test extrapolation methods.  In this work, we employ two different
approaches, one discussed in Sec.~\ref{sec:extrapolA} and the
other in Sec.~\ref{sec:extrapolB}. While we mainly emphasize the
ground state energy and point proton rms radius, we include
extrapolation of the ground state quadrupole moment and the
$3^+ \to 1^+$ E2 transition matrix element. The first set of
extrapolation methods is based on a combination of existing
techniques~\cite{Forssen:2008qp,Maris:2008ax,Coon:2012ab,Furnstahl:2012qg,More:2013rma}. The
second set is based on recent insights~\cite{Wendt:2015} on the
precise infrared (IR) length scale of the truncated many-body HO basis
that is employed in the NCSM and NCFC approaches. These two approaches
differ mainly in the precise definition of the IR length scale that is
being used, and in the inclusion of a phenomenological ultraviolet
(UV) correction in the first one. The large set of results that is
presented in this paper gives us an opportunity to compare the
perfomance of both sets of extrapolation methods.

First we introduce the elements appearing in the expressions for the
extrapolation functions.  Let $\lambda_t$ represent our IR regulator
that is implicit in the HO basis we have adopted.  This IR regulator
may be connected to an equivalent length of a box $L_t$, where
$L_t=1/\lambda_t$, as discussed in
Refs.~\cite{Furnstahl:2012qg,More:2013rma,Wendt:2015}. Furthermore, an
UV regulator $\Lambda$ may also be defined employing the duality of
position and momentum in the HO basis. We will return to the precise
definitions of these quantities in the following paragraphs.
\subsection{Extrapolation methods A5 and A3
\label{sec:extrapolA}}
In this set of methods, $L_t$ is defined through $L_t=L_2 + \Delta L$.  Here $L_2$
is a member of a class of length scales defined by $L_i = 1/\lambda_i$ with $\lambda_i = \sqrt{m\Omega/2(N_i+3/2)\hbar}$ while
$\Delta L=0.54437b(L_0/b)^{-1/3}$ and $b=\sqrt{\hbar/m\Omega}$.  We define $N_i$ as the maximum oscillator s.p.~quanta ($2n+l$) in the chosen basis with an added shift denoted by $i$ so that $N_i = max(2n+l) + i$.
The UV regulator $\Lambda$ is defined in connection with the properties of the highest-lying
s.p.~orbit in the chosen HO basis through $\Lambda_i=\sqrt{2(N_i+3/2)m\Omega/\hbar}$.
With our choice of $i=2$ and these definitions, we now define a 5-parameter function of the IR and UV
regulators ($\lambda_2,\Lambda_2$) for the ground state energy as
\begin{eqnarray}
&&E_{gs}=E_{\infty} + a e^{-c\Lambda^2_2} + E_{IR}(\lambda_t)\, ,\label{ExtrapE_1}\\
&&\,\,\,\,\,\,\,E_{IR}(\lambda_t)=d e^{-2k_\infty / \lambda_t}\, .\label{ExtrapE_2}
\end{eqnarray}
The 5 parameters ($E_{\infty}, a, c, d, k_\infty$) in Eqs.~(\ref{ExtrapE_1}) and (\ref{ExtrapE_2}) 
are determined by our fitting procedure which we define below.  In principle,
according to Refs.~\cite{Furnstahl:2012qg,More:2013rma} the
parameter $k_\infty$ may be determined by the least energy required to liberate a nucleon from
the nucleus.  In practice we determine it by fitting since the theoretical result for the
least energy required to liberate a nucleon is not known {\it a priori}.

With the form of the extrapolating function fixed in Eqs.~(\ref{ExtrapE_1}) and (\ref{ExtrapE_2}), we carry out a multi-step
procedure to extract the extrapolated (asymptotic) energy for the infinite Hamiltonian
matrix limit of the system.  We call this
extrapolation method ``A5" to represent its connection with ``Extrapolation A" of Ref.~\cite{Maris:2008ax} and to 
signify it has 5 free parameters.

Our multi-step fitting procedure for the ground state energy is as follows:
\begin{enumerate}
\item{Locate the minimum ground state energy with respect to $\hbar\Omega$
for the highest $N_{\rm max}$ employed}
\item{Take the value of $\hbar\Omega$ that is $2.5$~MeV above that minimum and four more $\hbar\Omega$ values
at additional increments of $2.5$~MeV}
\item{For this set of five $\hbar\Omega$ values, use five increments in $N_{\rm max}$ leading to the highest value
(excluding $N_{\rm max}=0$ so $N_{\rm max}=[2,4,6,8,10]$ is the lowest $N_{\rm max}$ set employed to generate an uncertainty as defined in step 6 below)}
\item{Weight each point using the finite difference from the previous $N_{\rm max}$ point with the exception of the first point that is weighted by the second weight scaled by the ratio of the second weight to the third weight}
\item{Perform a chi-square fit to these 25 points and obtain ($E_{\infty}, a, c, d, k_\infty$)}
\item{Assigned uncertainty is defined as 1/2 the spread in five separate extrapolates ($E_{\infty}$) obtained from
independent fits to the five $N_{\rm max}$ points at each of the five $\hbar\Omega$ values. One exceptional
case is $N_{\rm max}=[2,4,6,8]$ where we assign an uncertainty 1.5 times the uncertainty for
$N_{\rm max}=[2,4,6,8,10]$}
\item{Repeat for the next $N_{\rm max}=[4,6,8,10,12]$ set of 25 points, and with higher $N_{\rm max}$ sets as
far as they are available}
\item{Verify that a converging sequence of extrapolates is obtained that exhibits decreasing uncertainty with increasing
$N_{\rm max}$ and that successive extrapolates are consistent with previous extrapolates within the assigned uncertainties}.
\end{enumerate}

For the point proton rms radius we follow Refs.~\cite{Furnstahl:2012qg,More:2013rma} and adopt a 3-parameter function of the IR
regulator ($\lambda_2$) alone ({\it i.e.} there is no explicit dependence on the UV regulator) as

\begin{eqnarray}
r^2=r^2_{\infty}\left[1-(c_0\beta^3+c_1\beta)e^{-\beta} \right]\, .
\label{ExtrapR_1}
\end{eqnarray}
In Eq.~(\ref{ExtrapR_1}) the quantity $\beta = 2 k_{\infty} L_t$ with $k_{\infty}$ taken from theory as

\begin{eqnarray}
k_{\infty}=\frac{\sqrt{2m(E_s+E_c)}}{\hbar}
\label{kinf_thy}
\end{eqnarray}
where $E_c$ is the Coulomb barrier approximated as
$(Z-1)e^2/L_t$.  Here $e$ is the proton charge and Z = 3 for $^6$Li.
The quantity $E_s$ denotes the proton removal energy which we fix from the
extrapolated theoretical calculations using the same Hamiltonian.
For $^6$Li,
these proton removal energies (resulting a free space proton plus neutron plus $^4$He
since $^5$He is unbound) are calculated to be
3.25 (2.91)~MeV for JISP16 (NNLO$_{\rm opt}$) respectively.  These single-proton removal
energies take into account that the binding energy of $^4$He is 28.3 (27.6)~MeV
for JISP16 (NNLO$_{\rm opt}$) respectively and the extrapolated ground state energies
for $^6$Li obtained in the present work are 31.55 (30.51)~MeV for JISP16 (NNLO$_{\rm opt}$)
respectively.
Thus, the three parameters to fit are ($r^2_{\infty},c_0,c_1$).  We also attempted 4-parameter fits by
including $k_{\infty}$ among the fit parameters, but these fits revealed that these 4 parameters
were too many for the limited data range available.

Our multi-step fitting procedure for the point proton rms radius has important differences from the procedure
for the ground state energy and is as follows:
\begin{enumerate}
\item{Take results for $\hbar\Omega$ values in $2.5$~MeV increments covering $30$ to $40$~MeV}
\item{For this set of five $\hbar\Omega$ values, use five increments in $N_{\rm max}$ leading to the highest value
(excluding $N_{\rm max}=0$ so $N_{\rm max}=[2,4,6,8,10]$ is the lowest $N_{\rm max}$ set employed to generate an uncertainty as defined in step 5 below)}
\item{Weight each point using the finite difference from the previous $N_{\rm max}$ point with the exception of the first point that is weighted by the second weight scaled up by a factor of 3}
\item{Perform a chi-square fit to these 25 points and obtain ($r^2_{\infty},c_0,c_1$)}
\item{Assigned uncertainty is defined as 3/2 the spread in five separate extrapolates ($r^2_{\infty}$)
obtained from independent fits to 10 points from 5 $\hbar\Omega$ pairs \{(1,2),(2,3),(3,4),(4,5),(3,5)\}
where ``(1,2)'' represents the $\hbar\Omega$ pair $(30, 32.5)$~MeV, etc.}
\item{Repeat for the next $N_{\rm max}=[4,6,8,10,12]$ set of 25 points, and with higher $N_{\rm max}$ sets as
far as they are available}
\item{Verify that a converging sequence of extrapolates is obtained that exhibits decreasing uncertainty with increasing
$N_{\rm max}$ and that successive extrapolates are consistent with previous extrapolates within the assigned uncertainties}
\end{enumerate}
We call this extrapolation method for the rms radius ``A3".  

There are a number of ingredients to our fitting procedures, such as the assignments of weights and the binning of data
for the rms radii fits.  These choices were dictated by arguing that data closer to convergence are more significant and
by requiring consistency among the set of extrapolates.  We also tested these procedures on other light nuclei 
with available calculations and found they provide consistent results for these other systems.

\begin{figure}[htb]
\centering
\includegraphics[width=0.49\linewidth]{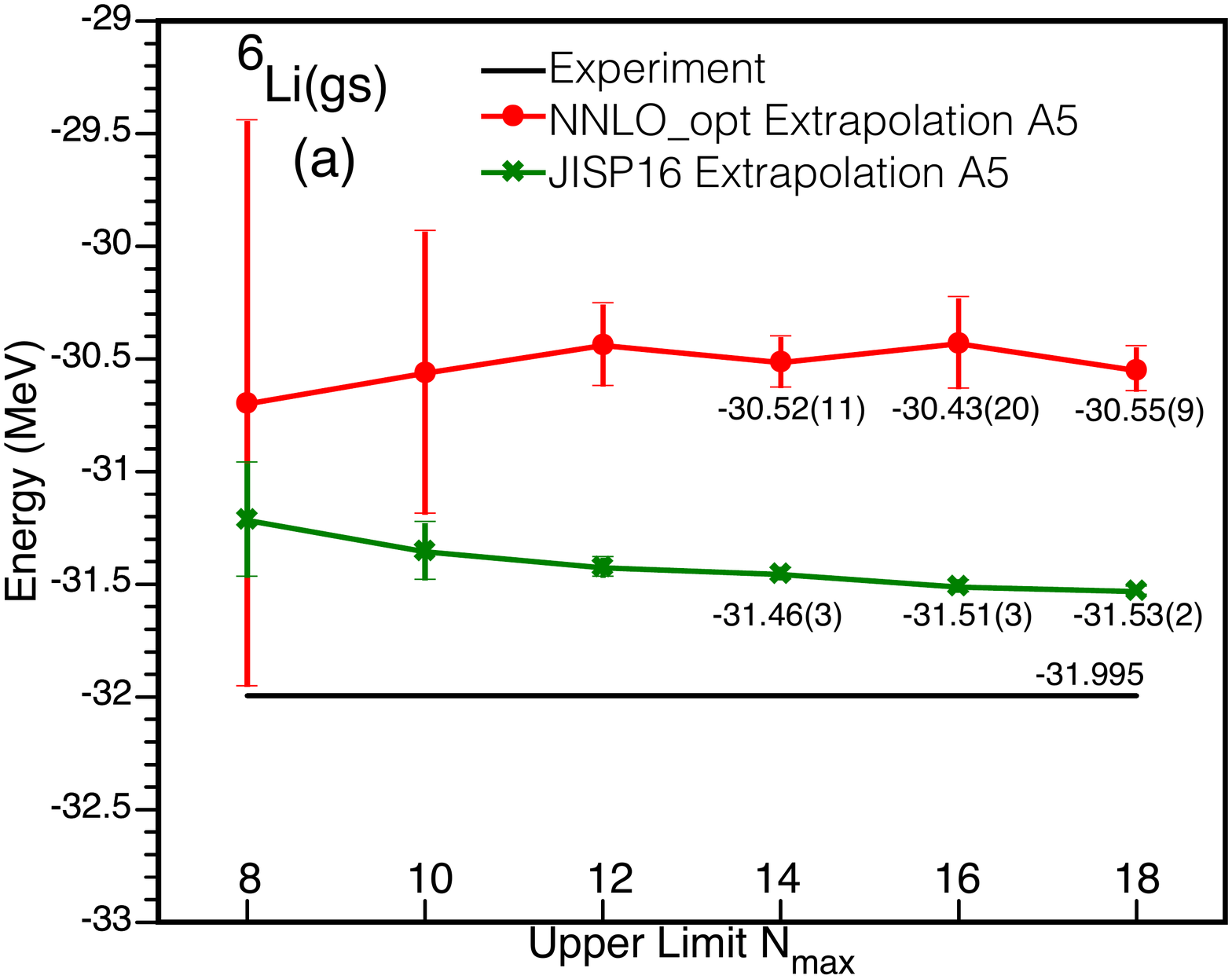}
\includegraphics[width=0.50\linewidth]{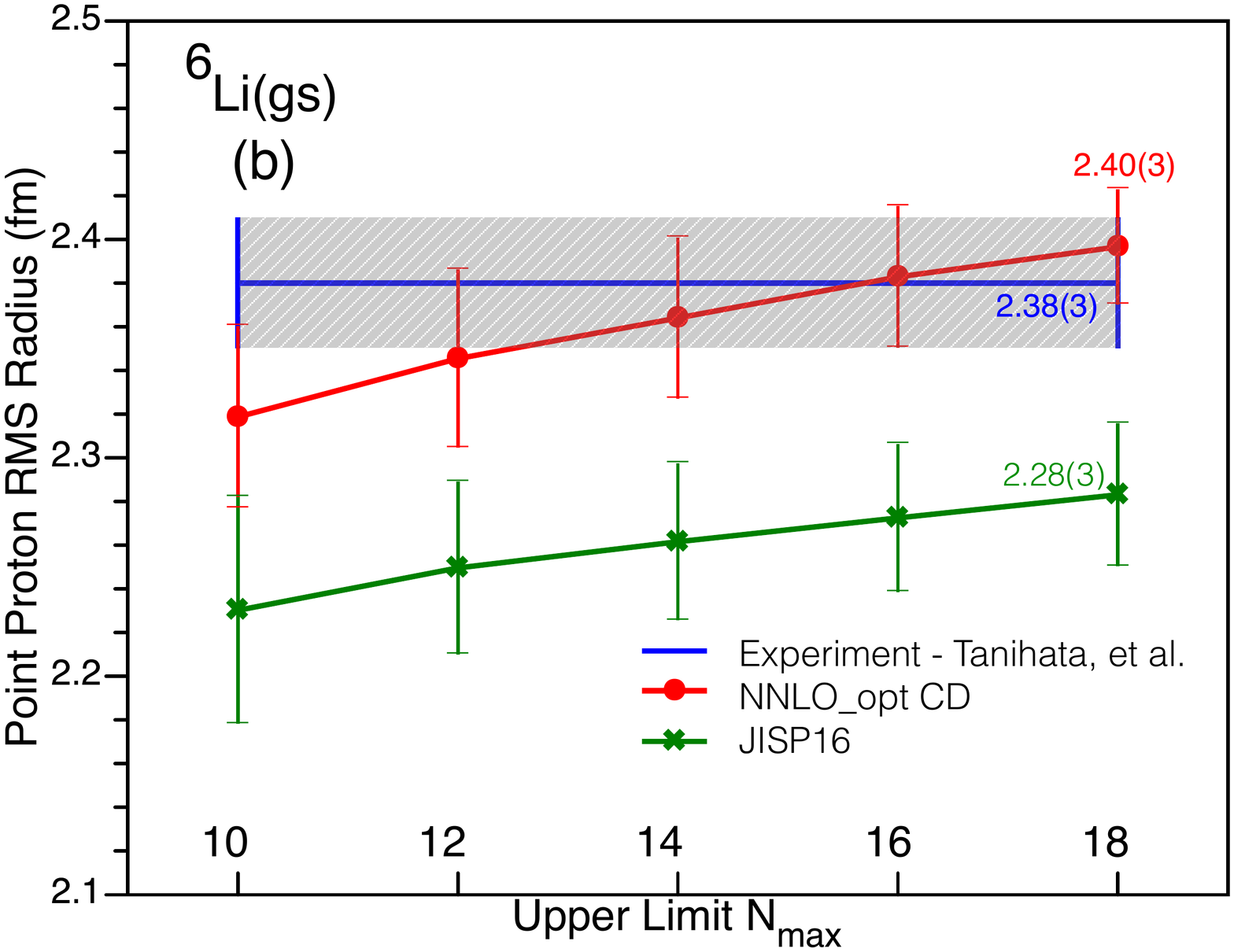}
\caption{(Color online) NCFC results for the ground state energy of $^6$Li (a) and the point proton rms
radius (b) using the JISP16 and NNLO$_{\rm opt}$ interactions.  The experimental results are also shown.
The extrapolation ``A5''  (``A3'') is employed for the ground state energy (point proton rms radius) as described in the text.  Uncertainties are indicated as error bars and are quantified using the rules specified in the respective procedures. Successive improvements are obtained by using finite matrix results for sets of
values of ($N_{\rm max}$, $\hbar\Omega$) governed by the rules with the highest value of $N_{\rm max}$ used to
place the result on the horizontal axis. Numerical results are quoted for selected cases along with the quantified uncertainty indicated in parenthesis as the amount of uncertainty in the least significant figures quoted.
\label{extraFig}}
\end{figure}

We employ these extrapolation methods for the ground state energy and the point proton rms radius of $^6$Li and show the results in Fig.~\ref{extraFig}.  Each discrete point in Fig.~\ref{extraFig} represents the result of applying the methods outlined above.   The horizontal axis defines the upper limit of $N_{\rm max}$ values used to determine the result plotted.   The defined uncertainty for each observable is depicted as an error bar for each point. That uncertainty for the points at $N_{\rm max}=10$ and above follow the scheme presented above.  For the exceptional points at $N_{\rm max}=8$ in the ground state energy figure, we have only 20 data points in the fits so we estimated the error as 1.5 times the error shown at $N_{\rm max}=10$.

For the ground state energies in Fig.~\ref{extraFig} we observe that the trend in the extrapolated energies is approximately flat with increasing $N_{\rm max}$ in the case of chiral NNLO$_{\rm opt}$ while there is a visible downward trend for JISP16.  For the point proton rms radii there is a systematic upward trend in the extrapolations.  This systematic upward trend indicates a possible additional contribution to the functional form
used in Eq.~(\ref{ExtrapR_1}).  The possibility of additional terms has been discussed in
Refs.~\cite{Furnstahl:2012qg,More:2013rma}.

Overall, there is approximate consistency in extrapolations shown in Fig.~\ref{extraFig} at successively
increasing values of $N_{\rm max}$ with the results at lower values of $N_{\rm max}$ as seen
by inspection of the quantified uncertainties.  For the ground state energies, those uncertainties are also
decreasing systematically with increasing $N_{\rm max}$.  However, for the point proton rms radius, the
uncertainties are only decreasing slowly with $N_{\rm max}$ over the range shown here.  This slow decrease
of the uncertainty with increasing $N_{\rm max}$ may serve as another indication of the
need for one or more additional terms in Eq.~(\ref{ExtrapR_1}).

\subsection{Extrapolation methods based on $L_\mathrm{eff}$
\label{sec:extrapolB}}
This set of methods is based on the recent work by
Wendt~\textit{et al.}~\cite{Wendt:2015}, that showed that a precise IR length scale
$L_\mathrm{eff}$ for many-body NCSM model spaces can
be determined by equating the intrinsic kinetic energy of $A$ nucleons
in the NCSM space to that of $A$ nucleons in a $3(A -1)$-dimensional
hyper-radial well with a Dirichlet boundary condition for the hyper
radius. In this approach, $L_\mathrm{eff}$ will be $A$-dependent and there
does not exist a closed-form expression but one has to rely on
numerical evaluation or on tabulated values. The corresponding UV
scale is defined as $\Lambda_\mathrm{eff} = L_\mathrm{eff} / b^2$,
where $b=\sqrt{\hbar / (m \Omega)}$ is the oscillator length.

The suggested extrapolation method in Ref.~\cite{Wendt:2015} relies on
data that has been obtained at sufficienly large $\Lambda_\mathrm{eff}$ so
that the computations are UV converged. The simple exponential form
\begin{equation}
E(L_\mathrm{eff}) = E_\infty + d e^{- 2 \kappa_\infty L_\mathrm{eff}}
\end{equation}
is then used for IR extrapolations of bound-state energies. This is
the same form as Eq.~\eqref{ExtrapE_1} but without the
phenomenological UV correction term. The ground state of $^6$Li,
calculated with NNLO$_{\rm opt}$, was studied already in
Ref.~\cite{Wendt:2015} and an extrapolated result
$E_\infty = -30.17$~MeV was obtained from data up to
$N_\mathrm{max} \le 14$. It was compared with the variational minimum
$E = -30.27$~MeV obtained at $N_\mathrm{max}=18$. This result is
verified by the present computations and serves as a benchmark of a
large-scale calculation with the two shell model codes.

\begin{figure}[htb]
\centering
\includegraphics[width=\linewidth]{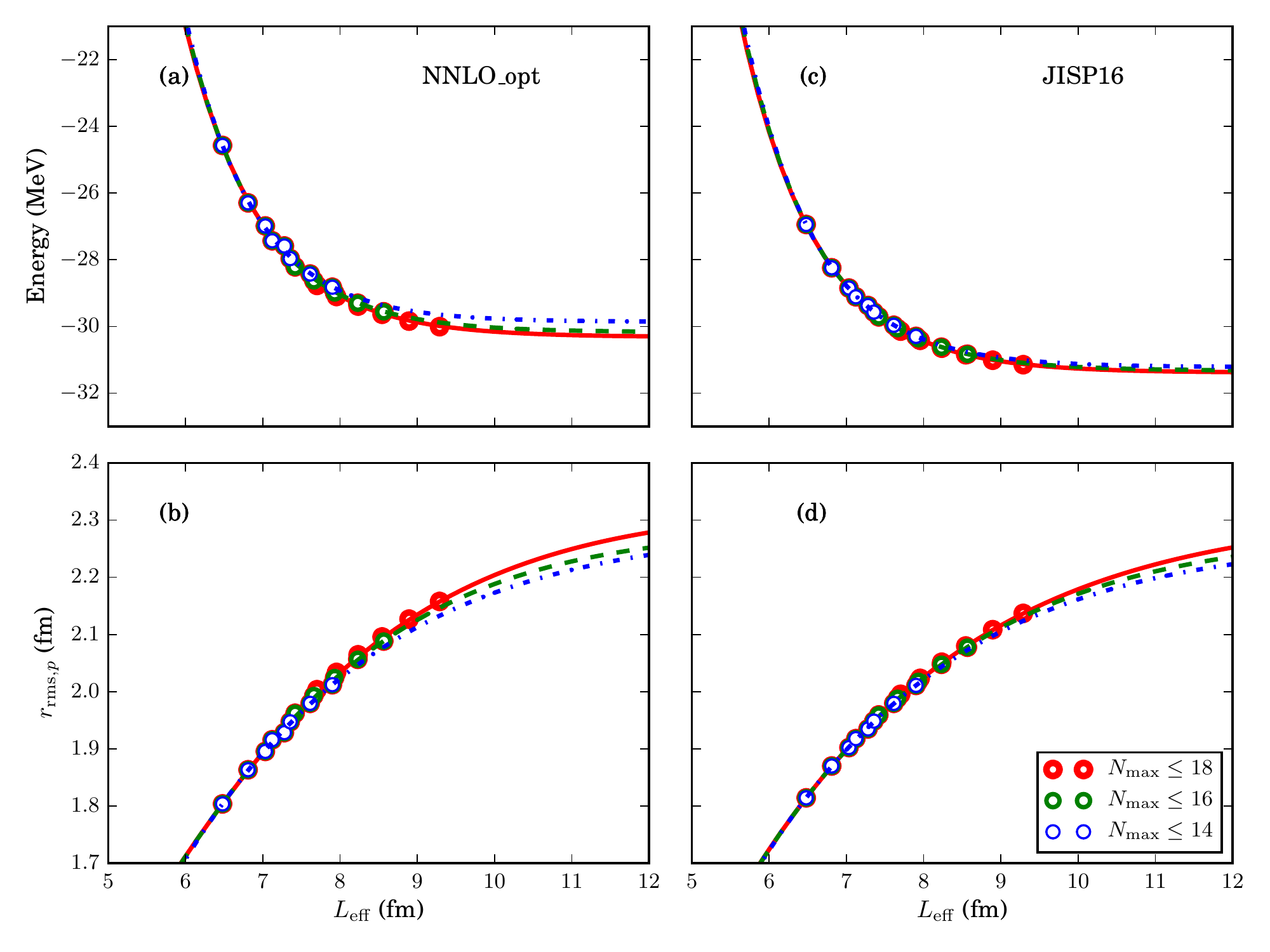}
\caption{(Color online) NCFC results for the ground state energy of
  $^6$Li (a,c) and the point proton rms
radius (b,d) using the JISP16 (right
panels) and NNLO$_{\rm opt}$ (left panels) interactions.
Extrapolations are performed in the IR length scale $L_\mathrm{eff}$
following the prescriptions of \cite{Wendt:2015, Furnstahl:2012qg}. Blue, green and red symbols (also with increasingly
thick linewidths) correspond to data with
$N_\mathrm{max} \le 14,16,18$ and the corresponding fits are shown
with dotted, dashed, and solid lines. All data is obtained with
$\Lambda_\mathrm{eff} \ge 1200$~MeV. \label{Leff-fig}}
\end{figure}

In this study we perform the extrapolation with all available data
($N_\mathrm{max} \le 18$) in order to get an NCFC result. We only use
data with $\Lambda_\mathrm{eff} \ge 1200$~MeV.  Truncations of the
data at $N_\mathrm{max} = 14$ and $N_\mathrm{max} = 16$ were also
performed to study the stability of the approach and to gauge one
aspect of systematic uncertainties. However, no attempt was made to
quantify systematic uncertainties from possible remaining UV
corrections. Results are shown in Figs.~\ref{Leff-fig}(a) and
\ref{Leff-fig}(c) for NNLO$_{\rm opt}$ and JISP16, respectively. This
approach gives the NCFC results $E = -30.32(7)$~MeV for
NNLO$_{\rm opt}$ and $E = -31.38(3)$~MeV for JISP16. The quoted
estimate of the uncertainly of the fit is obtained from refitting with
all possible pairs of data excluded from the data set.

In addition we performed extrapolations of the point proton rms radius of the $^6$Li ground state following the simple
functional form
\begin{equation}
r^2 = r_\infty^2 \left[ 1 - (c_0 \beta_\mathrm{eff}^3 + c_1
  \beta_\mathrm{eff}) e^{- \beta_\mathrm{eff}} \right], 
\label{eq:rp-leff}
\end{equation}
with $\beta_\mathrm{eff} = 2 \kappa_\infty L_\mathrm{eff}$. This
correction is from Ref.~\cite{Furnstahl:2012qg} but we use it here with
the correct IR scale $L_\mathrm{eff}$~\cite{Wendt:2015}. We note that
Eq.~\eqref{eq:rp-leff} is identical to Eq.~\eqref{ExtrapR_1} but that
here we allow $\kappa_\infty$ to be a free parameter.

We show results for point proton rms radii in Figs.~\ref{Leff-fig}(b) and
\ref{Leff-fig}(d) for NNLO$_{\rm opt}$ and JISP16, respectively. The
extrapolations using all data (always with
$\Lambda_\mathrm{eff} \ge 1200$~MeV) give $r = 2.32(9)$~fm for
NNLO$_{\rm opt}$ and $r = 2.31(6)$~fm for JISP16 with statistical
errors from the fit.

\begin{figure}[htb]
\centering
\includegraphics[width=\linewidth]{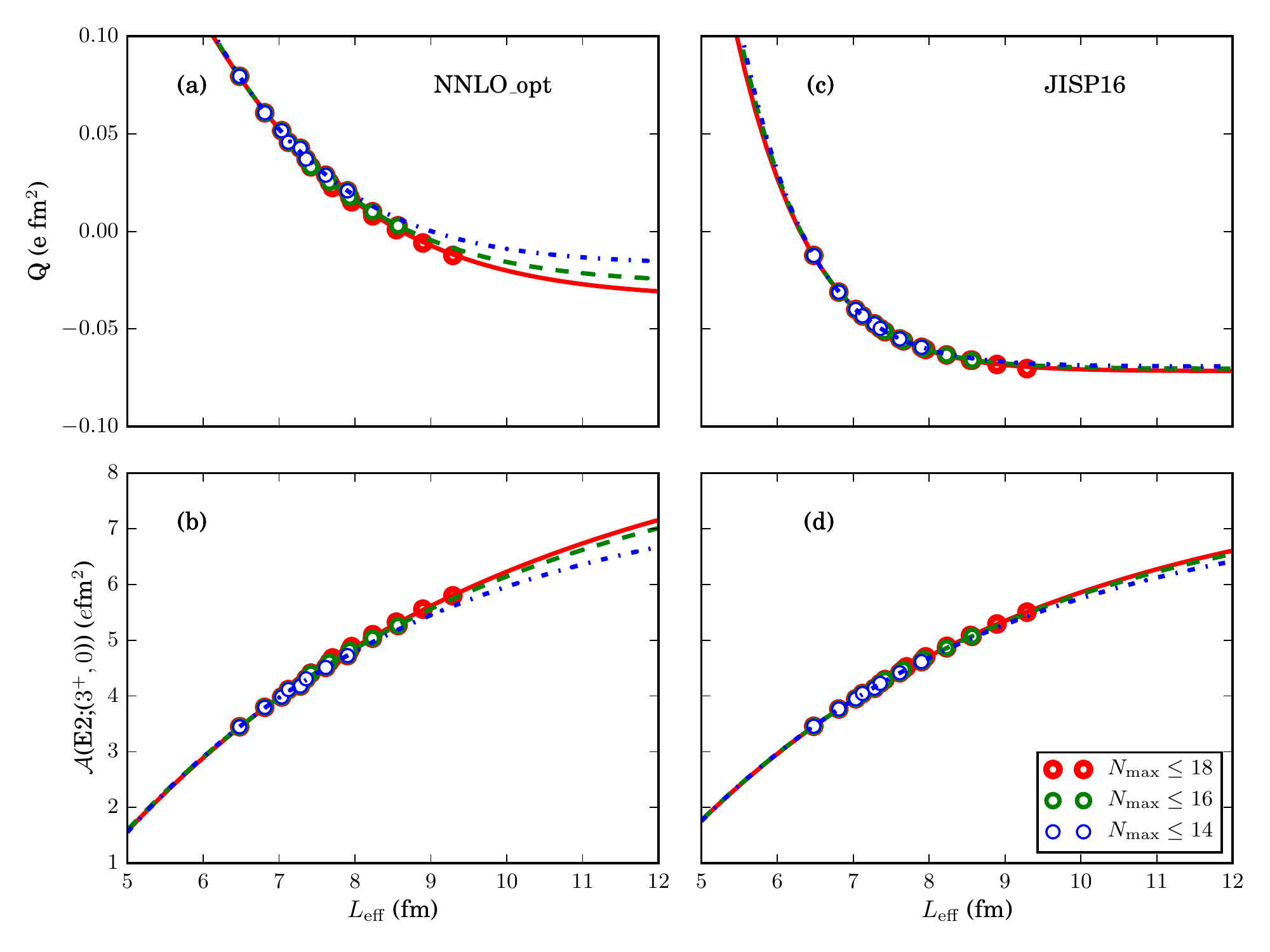}
\caption{(Color online) NCFC results for the quadrupole moment of
  $^6$Li($1^+$) (a,c), and the E2 matrix element for the $3^+ \to 1^+$
  transition (b,d) using the JISP16 (right panels) and NNLO$_{\rm opt}$
  (left panels) interactions.  Extrapolations are performed in the IR
  length scale $L_\mathrm{eff}$~\cite{Wendt:2015} following the
  prescriptions of \cite{Odell:2016}. Blue, green and red symbols
  (also with increasingly thick linewidths) correspond to data with
  $N_\mathrm{max} \le 14,16,18$ and the corresponding fits are shown
  with dotted, dashed, and solid lines. All data is obtained with
  $\Lambda_\mathrm{eff} \ge 1200$~MeV. \label{Leff-fig-E2}}
\end{figure}

The conclusions from this set of extrapolation methods are largely the
same as from extrapolations ``A5'' and ``A3''. While extrapolated
energies display a convincing stability, the extrapolated radii slowly
increase when data with higher $N_\mathrm{max}$ are included. This
hints to slow convergence and the need for subleading
corrections. Comparing results from two interactions it is clear
that the JISP16 results are better described by the leading-order
functional form than NNLO$_{\rm opt}$. This indicates that the former
interaction is softer and that computations are more UV converged.

Concerning the energy extrapolations we can conclude that the final
results differ by about 200 keV for the two extrapolation methods. It
seems that the use of a phenomenological UV correction makes a
detectable difference. Whether the extrapolated results are
consistent between the two approaches can only be determined once the
systematic uncertainties are available for both.
Our current results provide some very limited insights into the systematic uncertainties
since the $N_{\rm max}=18$ results for $^6$Li using JISP16 (NNLO$_{\rm opt}$)
have a minimum of $-31.43$~MeV ($-30.27$~MeV) at $\hbar\Omega=17.5$~MeV ($20$~MeV).
These upper bounds on the exact ground state energies are $50$~keV below ($50$~keV above)
the extrapolations using $L_\mathrm{eff}$ and $100$~keV above ($280$~keV above) the results
with extrapolation ``A5''. 

Finally, we performed extrapolations of the quadrupole moment of
the $^6$Li ground state and the $3^+ \to 1^+$ E2 transition matrix
element following the simple functional forms
\begin{align}
Q &= Q_\infty \left[ 1 - a \beta_\mathrm{eff}^3 e^{-
    \beta_\mathrm{eff}} \right], \\
\mathcal{A}_\mathrm{E2} &= \mathcal{A}_{\infty,\mathrm{E2}} \left[ 1 - a e^{-  \beta_\mathrm{eff}} \right]
\end{align}
with $\beta_\mathrm{eff} = 2 \kappa_\infty L_\mathrm{eff}$. These
corrections are from Ref.~\cite{Odell:2016}. 

We show results for the ground-state quadrupole moment in
Figs.~\ref{Leff-fig-E2}(a) and \ref{Leff-fig-E2}(c) for
NNLO$_{\rm opt}$ and JISP16, respectively. The corresponding results
for the E2 transition matrix element are in Figs.~\ref{Leff-fig-E2}(b)
and \ref{Leff-fig-E2}(d). 
The reduced transition strength is proportional to the transition
amplitude squared: $B(\mathrm{E2}) = (2 J_i +1)^{-1}
\mathcal{A}_\mathrm{E2}^2$, where $J_i$ is the total spin of the
initial state. 
The extrapolations using all data (always
with $\Lambda_\mathrm{eff} \ge 1200$~MeV) give $Q =
-0.034(3)$~$e
\mathrm{fm}^2$,
$B(\mathrm{E2};(3^+,0)) = 12(2)$~$e^2 \mathrm{fm}^4$ for
NNLO$_{\rm opt}$ and $Q = -0.072(1)$~$e \mathrm{fm}^2$,
$B(\mathrm{E2};(3^+,0)) = 9.2(6)$~$e^2 \mathrm{fm}^4$ for JISP16.
For all these cases, the quoted estimate of the uncertainly of the fit
is obtained from refitting with all possible pairs of data excluded
from the data set. This estimated uncertainty is quite large for the
E2 transition in particular.

\section{Ab initio NCFC results}
We now present our results for $^6$Li from NCFC calculations with both JISP16 and NNLO$_{\rm opt}$. We show the results for both natural and unnatural parity, and we employ finite matrix results up through
$N_{\rm max}=18$.  Our results are distinguished from those of Ref.~\cite{Cockrell:2012vd} in three ways:
We extend the upper limit of $N_{\rm max}$ (from 16 to 18); we include results for both parities;
and we include results for chiral NNLO$_{\rm opt}$.
Preliminary results of this study were presented in Ref.~\cite{Kim:2014iea}.

\begin{figure}[htb]
\centering
\includegraphics[width=1.0\linewidth]{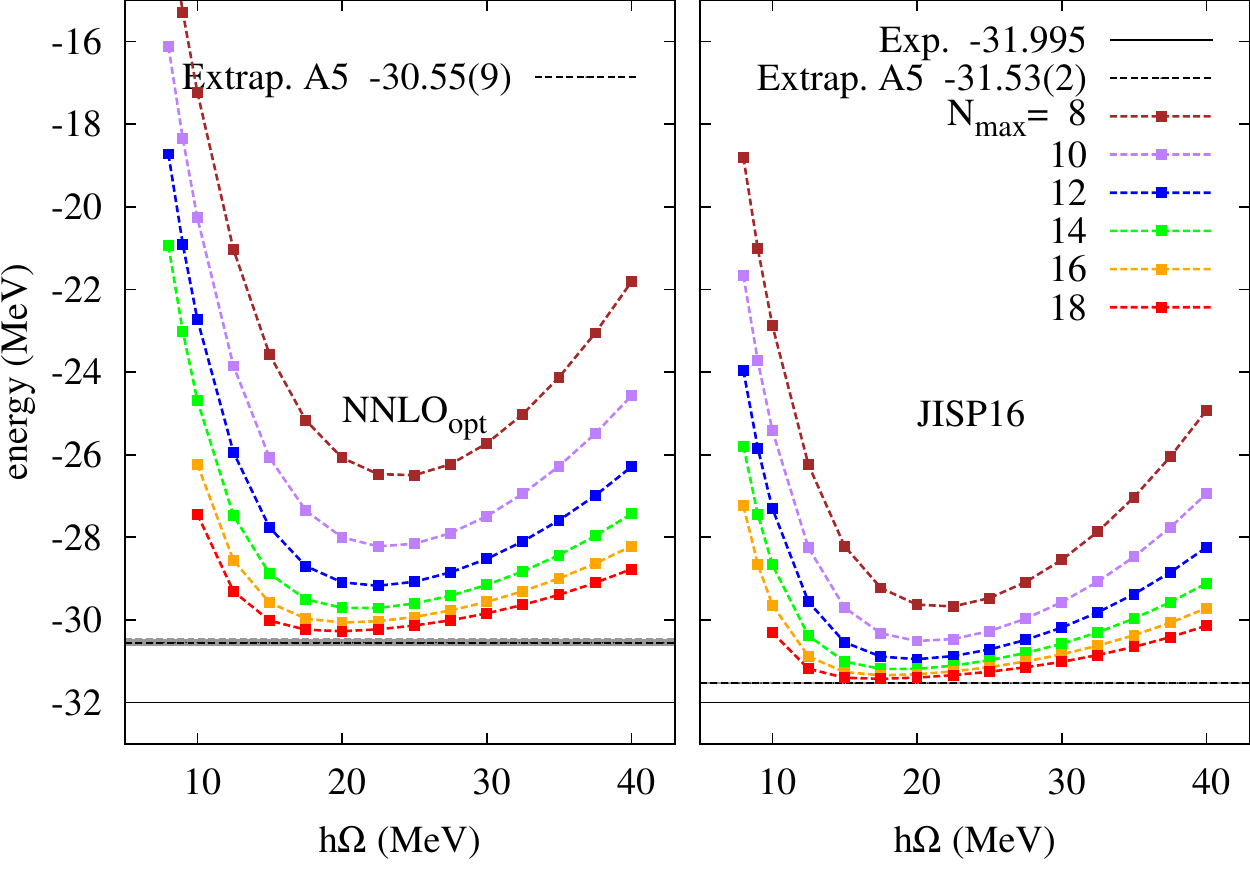}
\caption{(Color online)
The ground state energy of $^6$Li calculated with NNLO$_{\rm opt}$ and JISP16 as a function of the HO energy for
a range of $N_{\rm max}$ values. The result with the extrapolation A5 up to $N_\textrm{max}=18$ is indicated by the
black dashed line.
The shaded area around this extrapolation A5 result indicates our estimated uncertainty of $90$~keV for  NNLO$_{\rm opt}$
and $20$~keV for JISP16 respectively.}
\label{gs6Li}
\end{figure}

We present the ground state energy of  $^6$Li  as a function of the HO energy together with the result of  extrapolation A5 in Fig.~\ref{gs6Li} for both NN potentials.
As evident in Fig.~\ref{gs6Li} from the spread and compression of the U-shaped patterns,  the ground state energy using NNLO$_{\rm opt}$ converges more slowly than the ground state energy using JISP16.
The value obtained from extrapolation A5 using up to $N_{\rm max}=18$ with NNLO$_{\rm opt}$ indicates $^6$Li
is underbound by $1.44$~MeV. JISP16 also produces underbidding but only by $0.46$~MeV with extrapolation A5.

Both NNLO$_{\rm opt}$ and JISP16 are generating a net binding energy relative to their respective $\alpha$-$d$ thresholds. For NNLO$_{\rm opt}$ the calculated $\alpha$-$d$ threshold is $(-27.60)+(-2.22)=-29.82$~MeV.
For JISP16 the calculated $\alpha$-$d$ threshold is $(-28.30)+(-2.22)=-30.52$~MeV.
Thus, NNLO$_{\rm opt}$ (JISP16) binds $^6$Li at convergence relative to its respective $\alpha$-$d$ threshold by 0.73 (1.01)~MeV.

\begin{figure}[htb]
\centering
\includegraphics[width=1.0\linewidth]{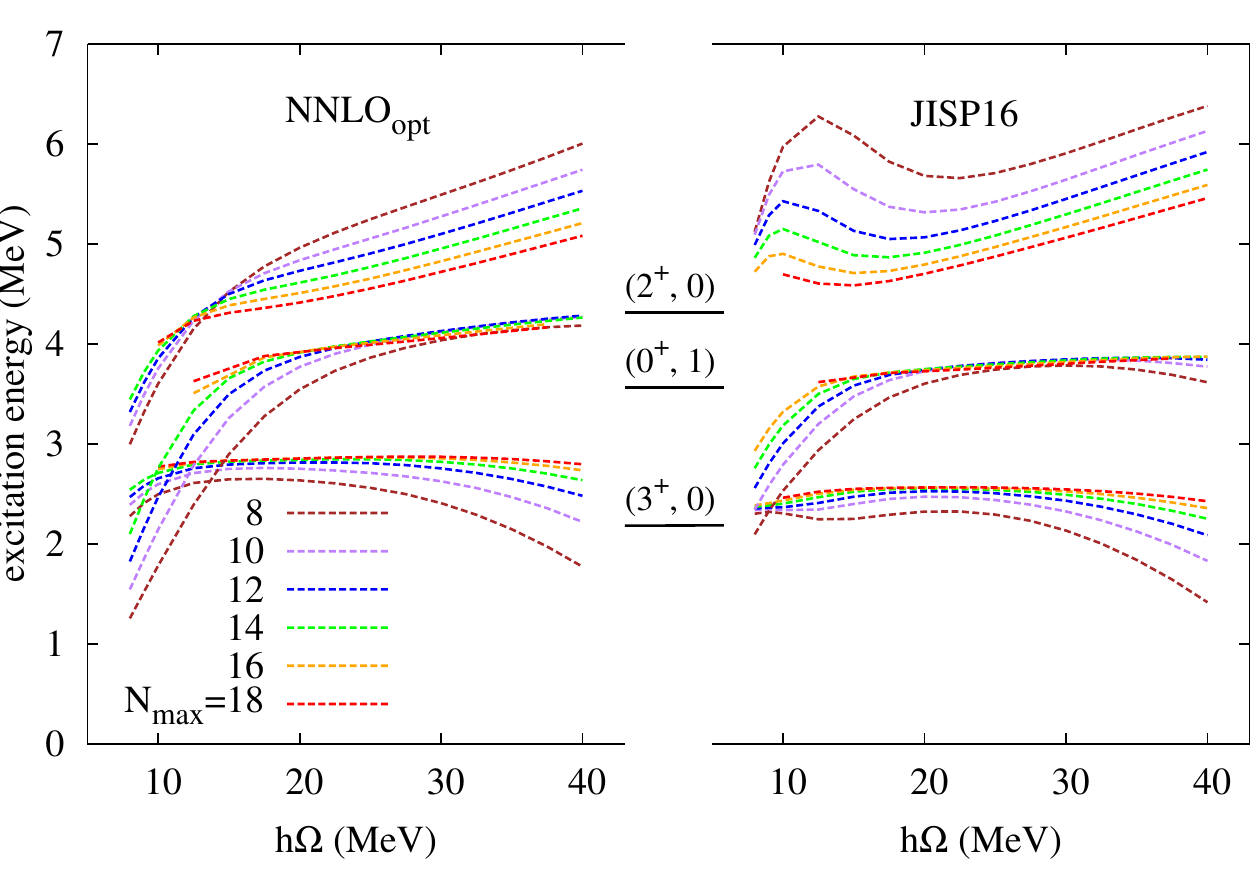}
\caption{(Color online)
Excitation energies for the three lowest natural parity excited states of $^6$Li calculated with (a) NNLO$_{\rm opt}$ and (b) JISP16 as a function of the HO energy up to $N_\textrm{max}=18$. The excitation energies for the three lowest experimental states are shown in the center. The order of these theoretical levels at the higher
$N_{\rm max}$ values corresponds to the experimental order over the full range of $\hbar\Omega$ depicted.}
\label{excitation6Li}
\end{figure}

The excitation energies relative to the  ground state energy through a sequence of $N_{\rm max}$ values are shown as functions of the HO energy in Fig.~\ref{excitation6Li} for both NNLO$_{\rm opt}$ and JISP16. The level ordering is the same for both interactions and in agreement with the experimental level ordering.  For the most part, the two interactions produce similar patterns as a function of $N_{\rm max}$ and $\hbar\Omega$.  For all three states the patterns tend towards convergence with increasing $N_{\rm max}$ and $\hbar\Omega$. That is, the curves trend towards a single curve and towards independence of $\hbar\Omega$ and
with increasing $N_{\rm max}$.

The exception to the similarity of the patterns for the two interactions is found for the $(2^+,0)$ excitation energy in the lower range of $\hbar\Omega$.  Here the NNLO$_{\rm opt}$ curves increase monotonically
with increasing $\hbar\Omega$ while the corresponding JISP16 curves display a peak
in the lower $\hbar\Omega$ range.  The peak
structure diminishes with increasing $N_{\rm max}$ so the curve for $(2^+,0)$ state with JISP16 grows increasingly similar to the corresponding curve with NNLO$_{\rm opt}$.

Let us focus on the excitation energies at $N_{\rm max}=18$ in Fig.~\ref{excitation6Li}.  We note that all
three excited states are above their interaction-specific $\alpha$-$d$ thresholds (results quoted above).
Therefore, these are continuum states and the finite HO matrix representation is providing a
bound state description of these resonances.  The connection between the HO basis description
as a bound state and the J-matrix description of continuum states in light nuclei has been extensively investigated
in Refs.~\cite{Shirokov:2008jv_prelim,Shirokov:2008jv,Mazur:2014_NTSE}.  For our purposes, it is important
to note that these investigations have shown a strong correlation between the $\hbar\Omega$-dependence of
the eigenvalue in the HO basis at fixed $N_{\rm max}$ for a state in the continuum with its resonance width:
stronger $\hbar\Omega$-dependence is indicative of a larger width.  At this time, this provides a qualitative indication how our observed $\hbar\Omega$-dependence would translate into relative widths of our states.
For an introduction to one method of calculating the resonance widths directly with continuum techniques, 
we defer to our discussion below of the Gamow Shell Model applied to $^6$Li.

The slopes in $\hbar\Omega$ of the three lowest excited states in $^6$Li shown in
Fig.~\ref{excitation6Li} suggest that, for both NNLO$_{\rm opt}$ and JISP16, there is
a progression from a relatively narrow $(3^+,0)$ to an intermediate $(0^+,1)$ followed
by a broad  $(2^+,0)$.
Experimentally, the $(3^+,0)$ has a width of $24$~keV.  The $(0^+,1)$ is an isobaric analog
state with a very narrow resonance width of $8.2$~eV, a property expected for isobaric analog states.
On the other hand,
the experimental width of the $(2^+,0)$ state is about $1.3$~MeV.
Therefore, the $N_{\rm max}$-dependence in the HO basis calculations with both interactions
reflects the experimental trends, especially when one considers the isobaric analog
nature of the $(0^+,1)$ state.

\begin{figure}[htb]
\centering
\includegraphics[width=0.72\linewidth]{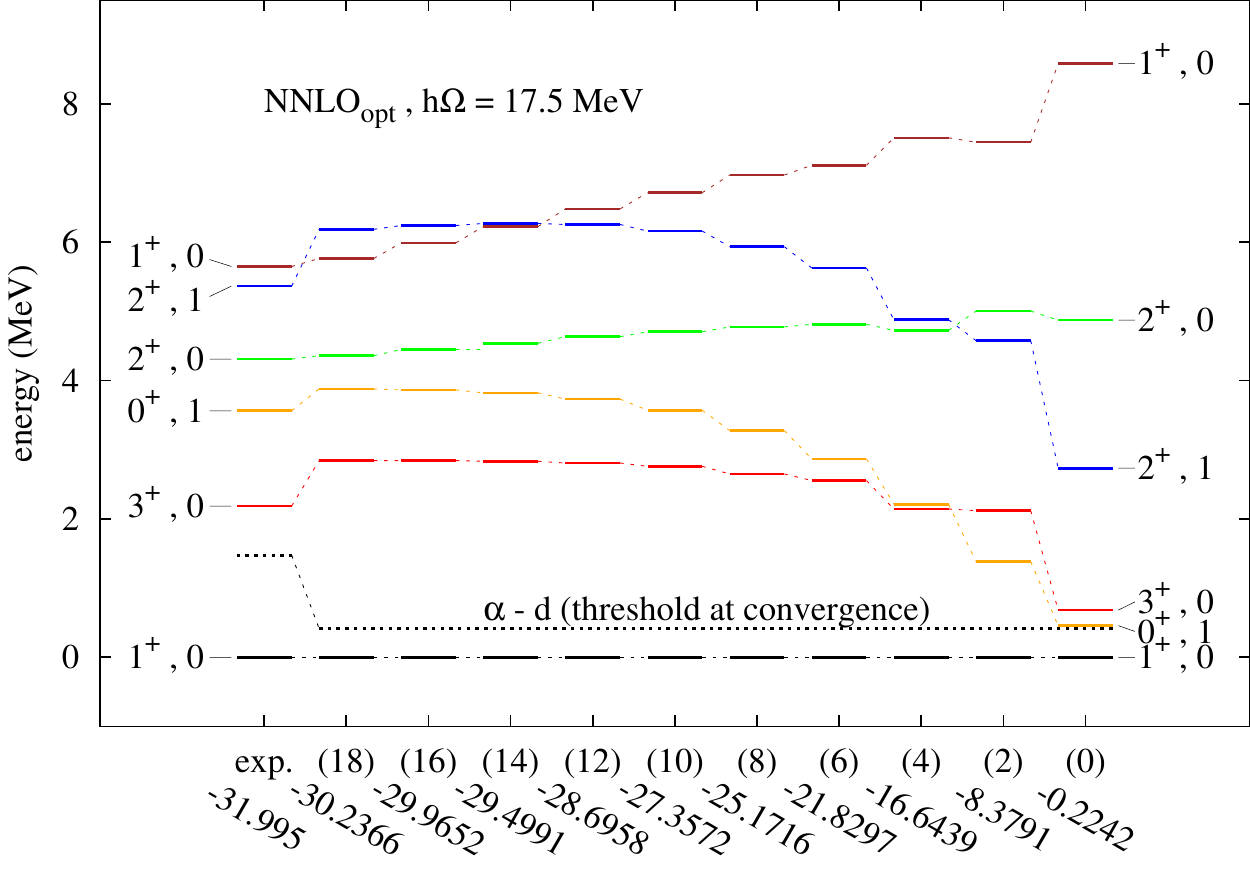}
\includegraphics[width=0.72\linewidth]{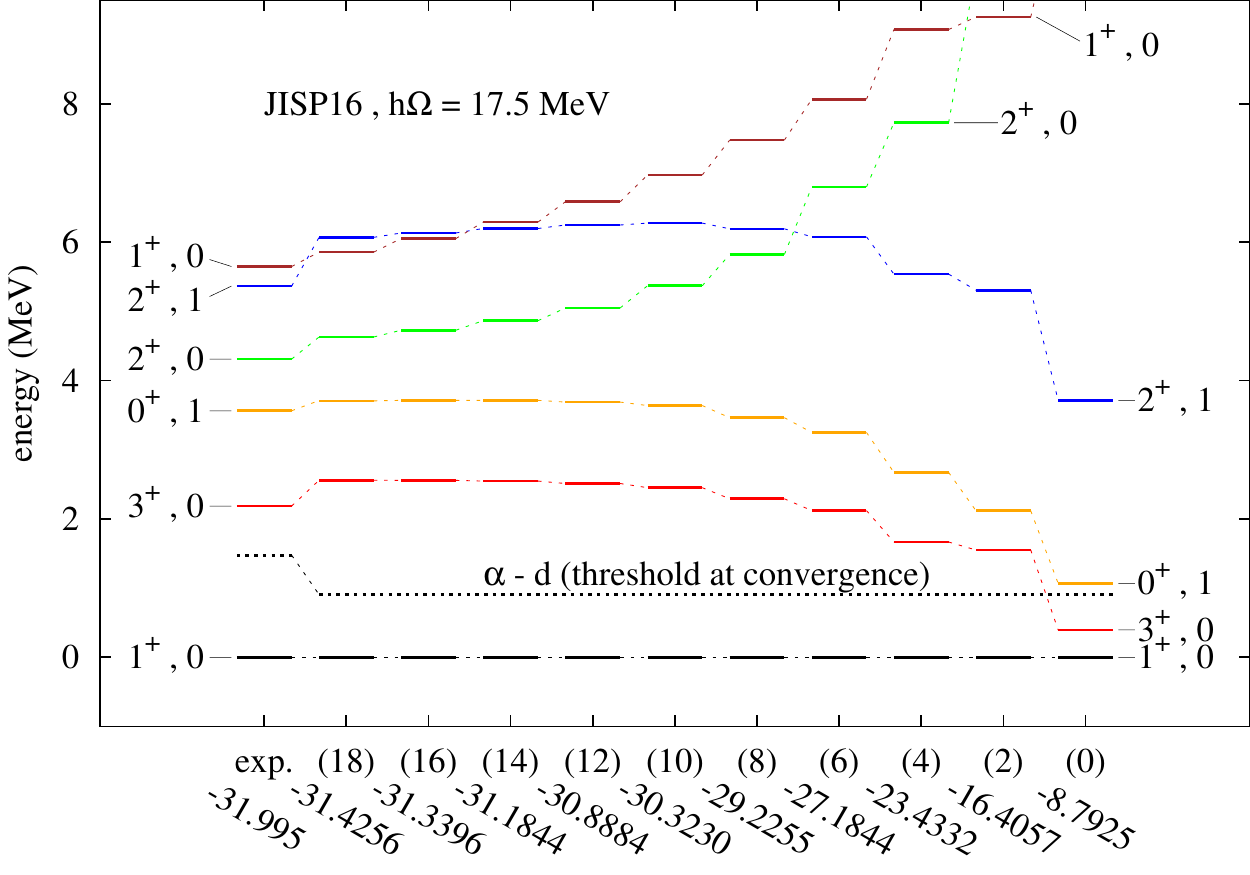}
\caption{ (Color online)
Excitation energies of the $^6$Li natural parity states calculated with NNLO$_{\rm opt}$ (top) and JISP16 (bottom) compared with experimental data. The calculated $\alpha$-$d$ threshold at convergence for each interaction is also depicted indicating all calculated excited states are in the continuum.
The calculated spectra are shown as a function of $N_{\rm max}$ which is indicated in parenthesis
below each column.
The ground state eigenvalue (in MeV) is also listed for each $N_{\rm max}$ below the column.}
\label{natural}
\end{figure}

We present in Fig.~\ref{natural} the excitation spectrum for both NNLO$_{\rm opt}$ (top) and JISP16 (bottom)
as a function of $N_{\rm max}$. For Fig.~\ref{natural} we take $\hbar\Omega =17.5$~MeV
which approximately provides the minimum value of the ground state energy at larger $N_{\rm max}$.
All states display a flow pattern towards convergence with increasing $N_{\rm max}$.  There is reasonable
agreement between theory and experiment at the highest values of $N_{\rm max}$.
The fourth and fifth excited states have
an inverted order compared with experiment for both NNLO$_{\rm opt}$ and JISP16.
Overall, the theoretical spectra appear to be converging to spectra that are spread wider in energy than the
experimental spectra.  This is a rather common feature of spectra obtained in
{\it ab initio} NCSM and NCFC calculations.

\begin{figure}[htb]
\centering
\includegraphics[width=0.72\linewidth]{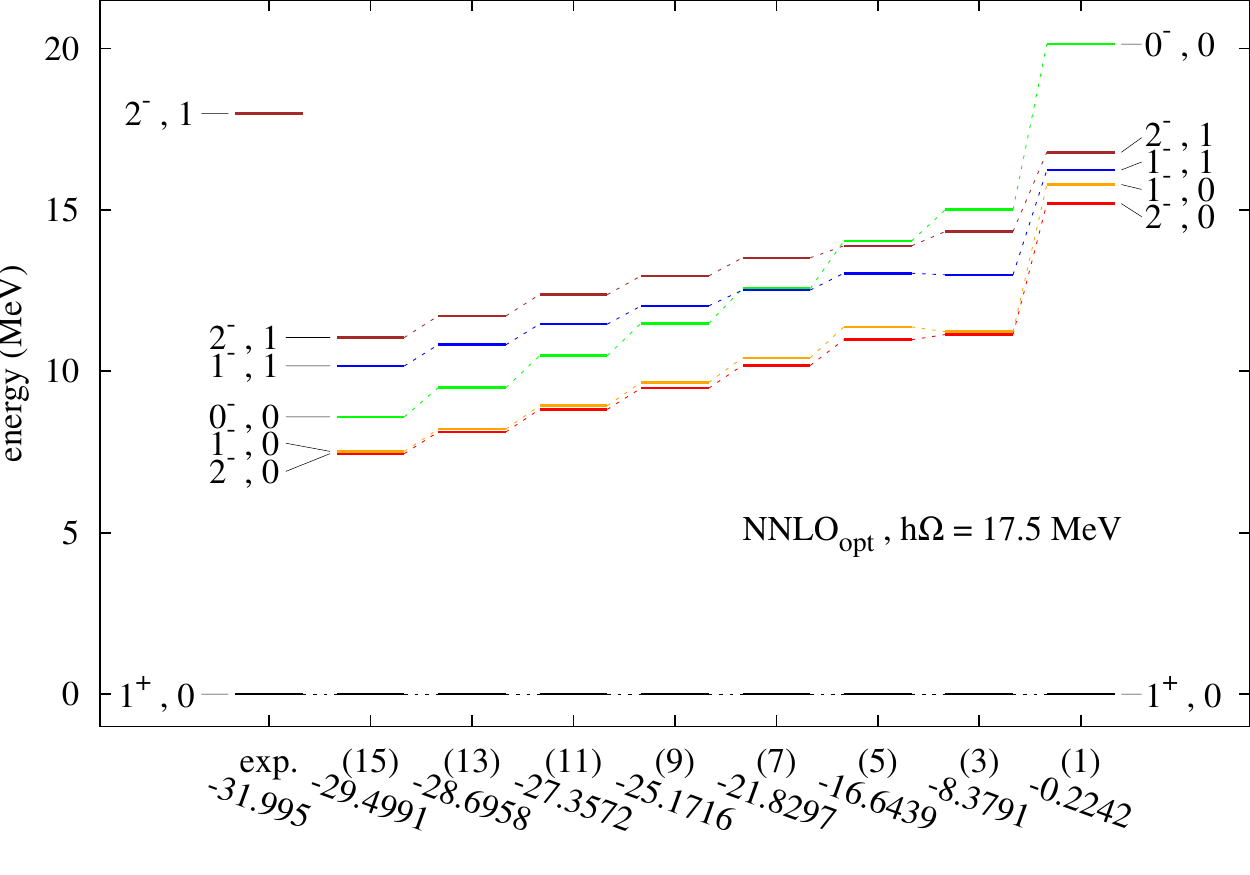}
\includegraphics[width=0.72\linewidth]{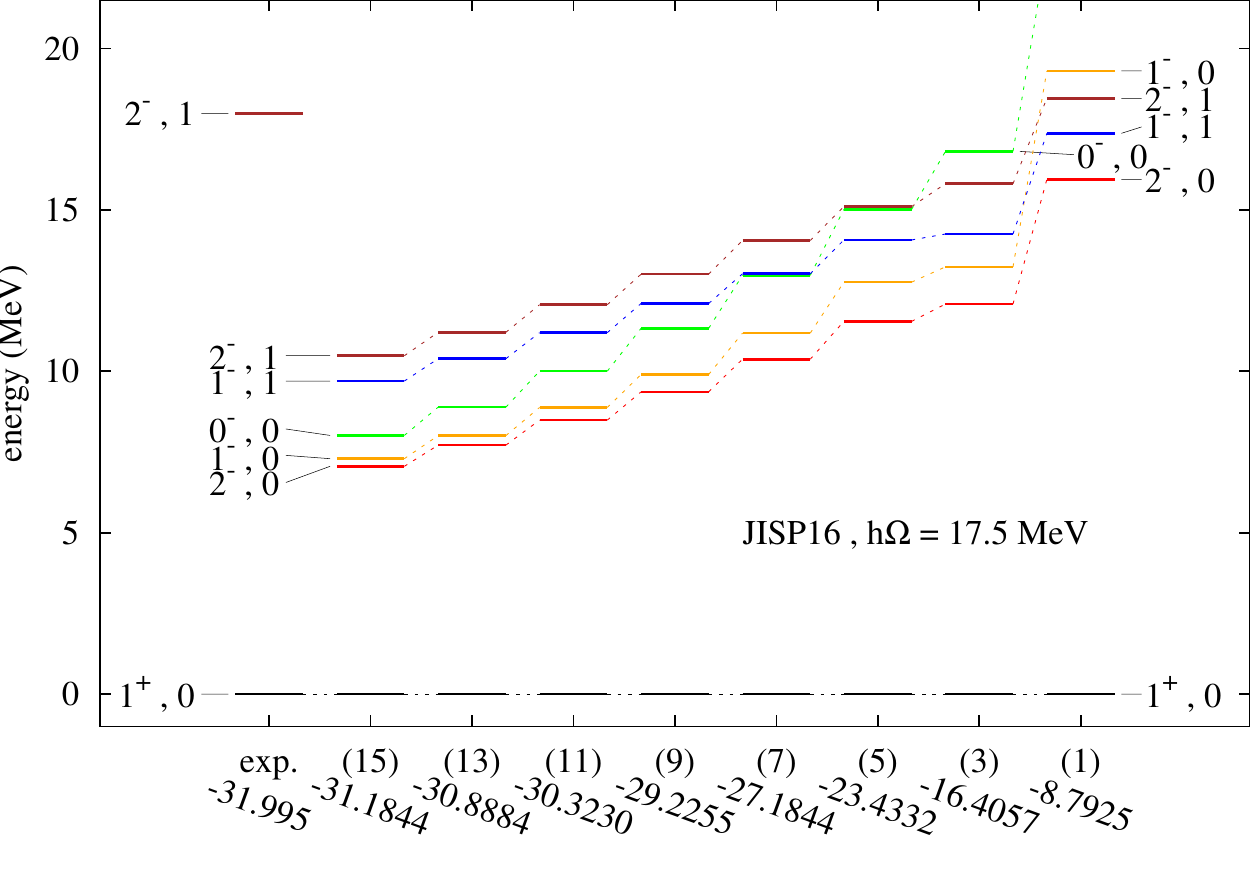}
\caption{ (Color online)
Excitation energies of $^6$Li unnatural parity states calculated with NNLO$_{\rm opt}$ (top) and JISP16 (bottom).
Here only a $(2^-,1)$ state is known experimentally below $20$~MeV and appears at $17.98$~MeV.
The calculated spectra are shown as a function of $N_{\rm max}$ which is indicated in parenthesis
below each column.
The ground state eigenvalue (in MeV) is also listed for each $N_{\rm max}-1$ and is the value used
to fix the excitation energy of these states at $N_{\rm max}$.}
\label{unnatural}
\end{figure}

Before discussing additional observables for the natural parity states, we introduce our solutions for
the unnatural parity states.
We solve for the lowest five eigenvalues at odd integer values of $N_{\rm max}$ and present
the results as excitation energies in Fig.~\ref{unnatural}.
The excitation energies of unnatural parity states at each $N_{\rm max}$ are defined relative to the ground state at $N_{\rm max}-1$.
In experiments, the lowest unnatural parity state is $(2^-,1)$ which is a broad resonance with a width of
$3$~MeV.
We observe that the energies of these calculated unnatural parity states are decreasing monotonically
with $N_{\rm max}$.  This pattern suggests that these states belong to the non-resonant continuum
of  $^6$Li.  Based on these results, we may infer that unnatural parity resonances in $^6$Li are likely to be
above $10$~MeV of excitation, which is consistent with available experimental information.

We note both the similarity of the level orderings and the similarity of the flow of the spectra between two
interactions shown in Fig.~\ref{unnatural}.  This suggests that the contributions of the negative
parity continuum to, for example, $d+\alpha$ scattering cross sections will be similar with these two
interactions.

\begin{figure}[htb]
\centering
\includegraphics[width=0.9\linewidth]{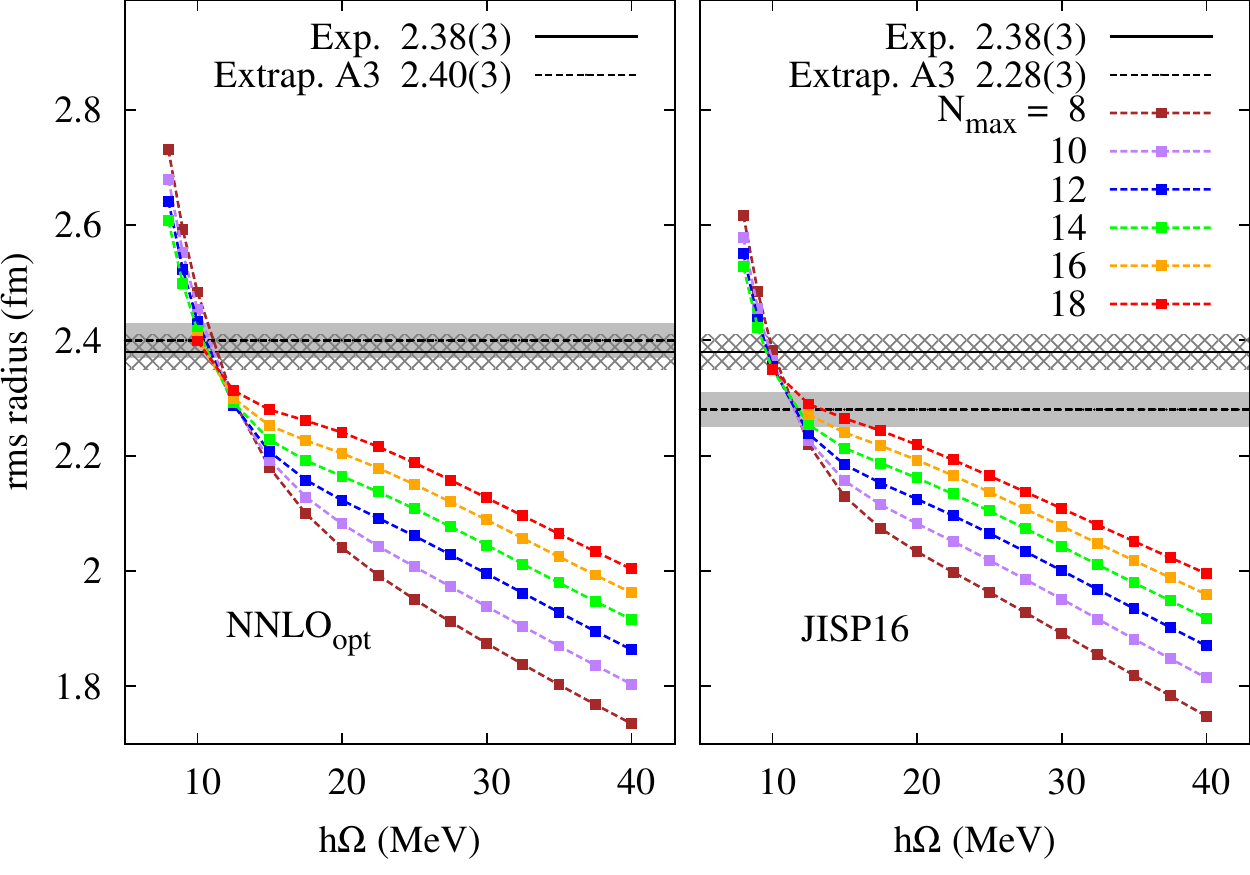}
\caption{(Color online)
Point proton rms radius of the ground state of $^6$Li as a function of HO energy for a sequence of $N_{\rm max}$ values.
The extrapolation A3 as discussed in the text is used in this case.
Uncertainties in the experimental values are indicated with a cross-hatch pattern while estimates of the uncertainty in the theoretical values are indicated by a grey band.
The experimental value is given by Ref.~\cite{Tanihata:2013jwa}.}
\label{ppRMSradius}
\end{figure}

We now return to the discussion of additional observables for the natural parity states.
Fig.~\ref{ppRMSradius} presents the point proton rms radius of the ground state of $^6$Li
for both interactions as a function of $\hbar\Omega$. Here we take $N_{\rm max}=8\sim 18$.
We observe a pattern similar to those seen in other NCSM investigations
and with other interactions~\cite{Caurier:2005rb,Barrett:2013nh} and/or other s.p.~basis spaces~\cite{Caprio:2012rv}.
It is also worth emphasizing that our calculated point proton rms radii are defined with respect to
the CM of the nucleus and are, therefore, free of spurious CM motion contributions.

We employ here the extrapolation A3 discussed in the previous section to obtain the infinite matrix limit
of the point proton rms radius along with a quantified uncertainty for each interaction.
The extrapolated rms radii shown in Fig.~\ref{ppRMSradius} are based on the $N_{\rm max}=[10,12,14,16,18]$
results (see Fig.~\ref{extraFig}).
We comment that the extrapolated point proton rms radius is close to the experimental value for
NNLO$_{\rm opt}$, while it is somewhat smaller than the experimental value for JISP16 interaction.
However, we do not emphasize the proximity of the theoretical and experimental point proton rms radii
for two reasons.  First, the convergence patterns and quantified uncertainties shown in Fig.~\ref{extraFig} indicate
we are still quite far from the converged result.  Second, these same results in Fig.~\ref{extraFig} also
indicate we are likely to have a systematic uncertainty in the extrapolation function defined in Eq.~(\ref{ExtrapR_1}) that is separate from our quantified uncertainty.

\begin{figure}[htb]
\centering
\includegraphics[width=0.9\linewidth]{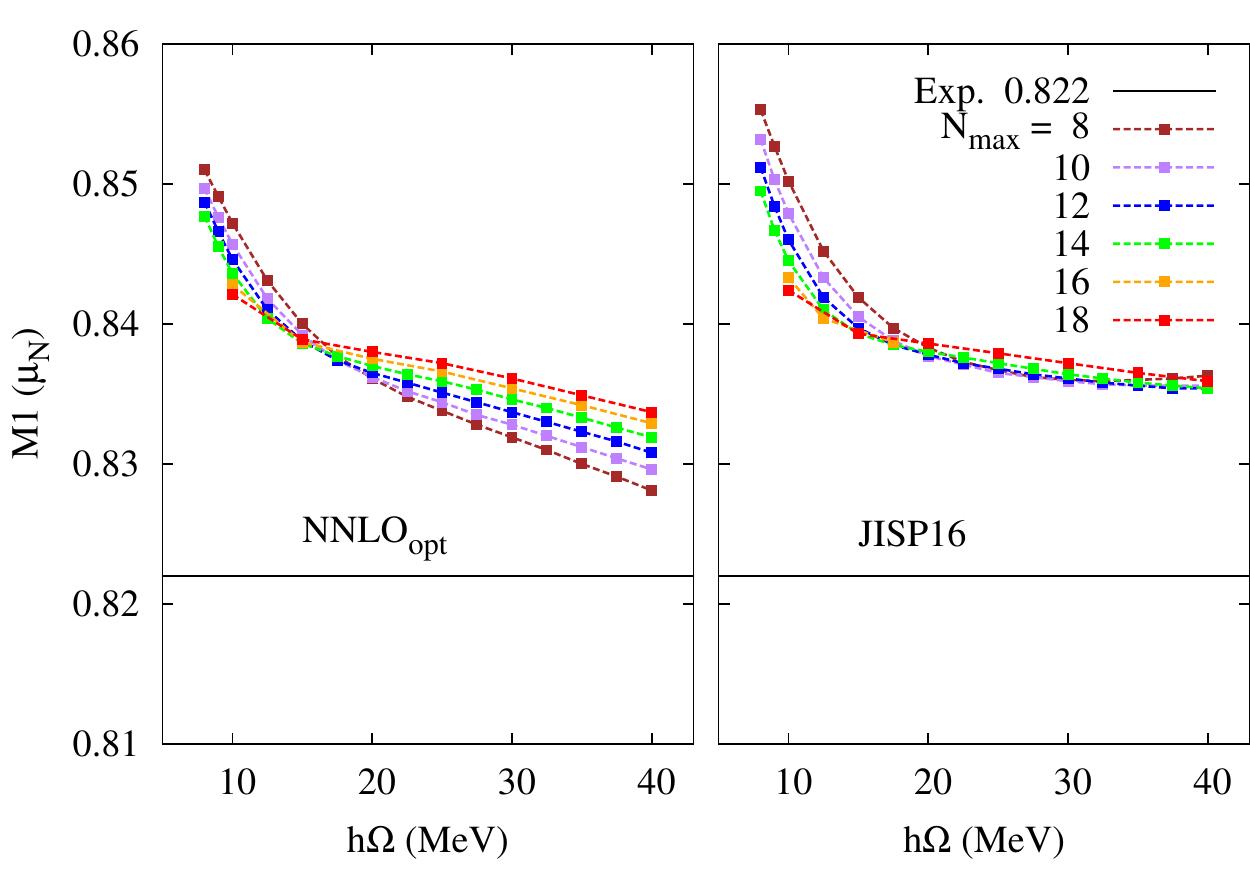}
\caption{(Color online) Magnetic dipole moment of the ground state of $^6$Li as a function of the HO energy
over a range of $N_{\rm max}$ values for the NNLO$_{\rm opt}$ (left panel) and the 
JISP16 (right panel) interactions.
Note the expanded scale.
The experimental value is from Refs.~\cite{AjzenbergSelove:1988ec,Tilley:2002vg}.}
\label{magneticDipole}
\end{figure}

We show the  magnetic dipole moment of the  ground state as a function of $\hbar\Omega$ with $N_{\rm max}$ values $8\sim 18$ in Fig.~\ref{magneticDipole}.
The results from both interactions appear to be converging to a result within 2\% of the experimental value.
A closer look reveals that the magnetic moment with JISP16 converges somewhat better than the result with NNLO$_{\rm opt}$.
Note that we have not evaluated theoretical corrections to the magnetic dipole operator which are needed
to making precise comparisons between theory and experiment.

\begin{figure}[htb]
\centering
\includegraphics[width=0.9\linewidth]{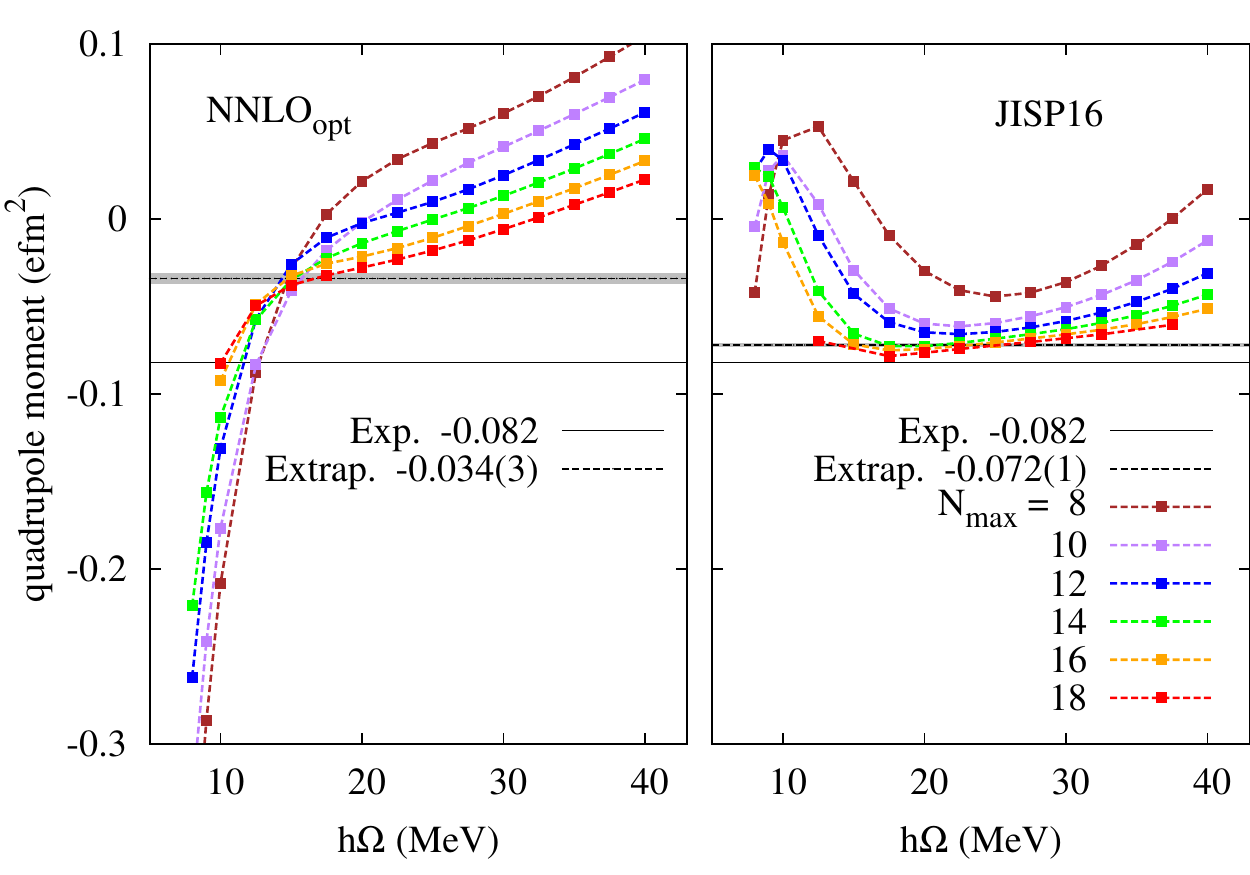}
\caption{(Color online) Quadrupole moment of the ground state as a function of the 
HO energy for a sequence of $N_{\rm max}$ values.
The extrapolation as discussed in the Sec.~\ref{sec:extrapolB} is used.
Estimates of the uncertainty are indicated by a grey band.
The experimental value is from Refs.~\cite{AjzenbergSelove:1988ec,Tilley:2002vg}.}
\label{quadrupole}
\end{figure}

The ground state quadrupole moment of $^6$Li represents a sensitive test of the 
wave function since delicate cancellations between positive and negative contributions 
are needed to explain the nearly vanishing
experimental value as explained in Ref.~\cite{Cockrell:2012vd}.
Fig.~\ref{quadrupole} displays our results for this ground state quadrupole moment for both interactions as a function of $\hbar\Omega$ for $N_{\rm max}$ values $8\sim 18$. Both interactions show trends favorable to agreement with experiment in the infinite matrix limit.
However, the convergence patterns are strikingly different from each other.
The pattern with NNLO$_{\rm opt}$ in Fig.~\ref{quadrupole} is reminiscent of the pattern seen in Fig.~\ref{magneticDipole} for the ground state magnetic dipole moment for NNLO$_{\rm opt}$ but inverted in
the ordering of the results with increasing $N_{\rm max}$.
On the other hand, the pattern with JISP16 in Fig.~\ref{quadrupole} is more like the pattern for the ground state energy seen in Fig.~\ref{gs6Li} except at low $\hbar\Omega$ values.
These two different patterns for the ground state quadrupole moment represent one of the more distinctive differences in the calculated results that we obtain for NNLO$_{\rm opt}$ and JISP16.  As in the case
of the magnetic moment, the quadrupole moment with JISP16 seems to be converging somewhat faster than the quadrupole moment with NNLO$_{\rm opt}$.

\begin{figure}[htb]
\centering
\includegraphics[width=0.75\linewidth]{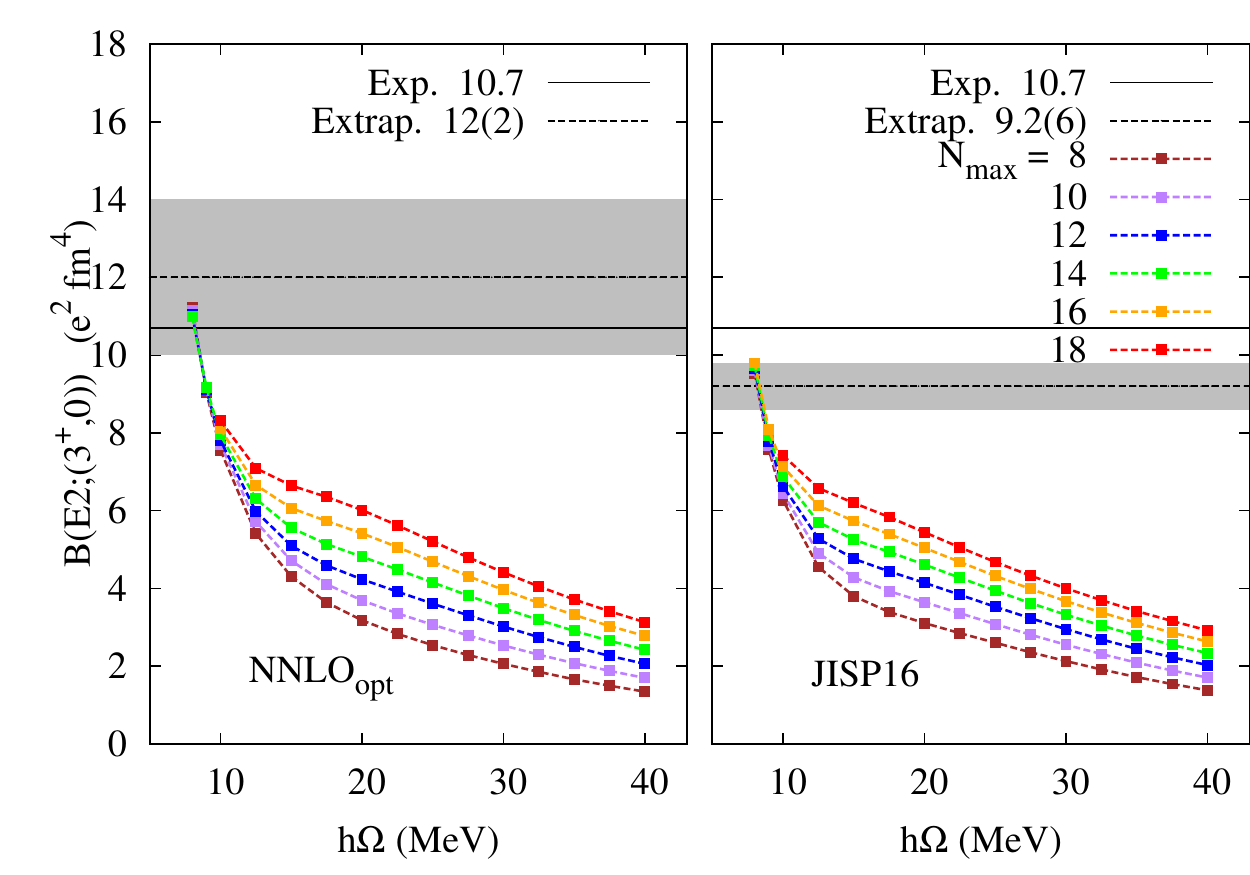}
\includegraphics[width=0.75\linewidth]{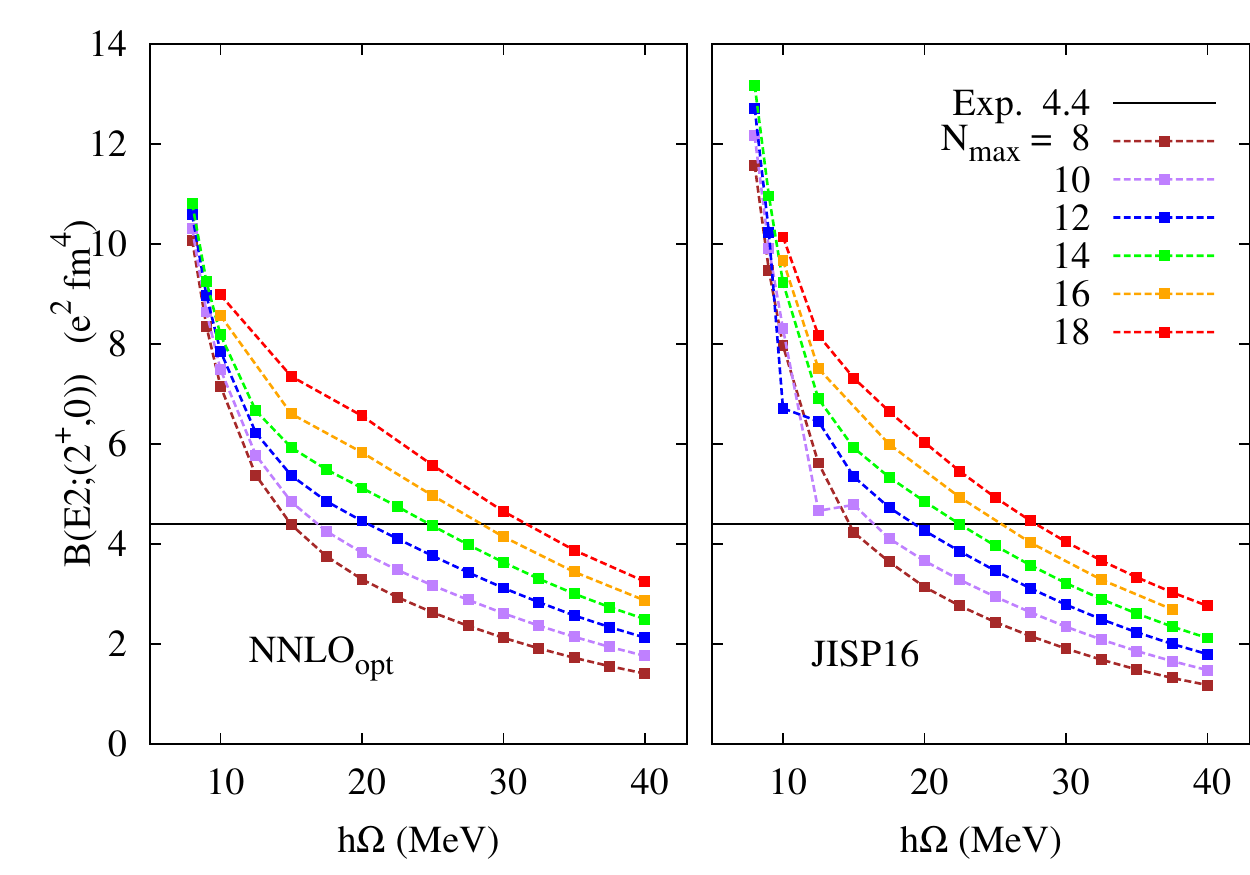}
\caption{(Color online) B(E2) to the ground state $(1^+,0)$ from $(3^+,0)$ (top) and $(2^+,0)$ (bottom) as a function of the 
HO energy for a sequence of $N_{\rm max}$ values.
The extrapolation of top panels are discussed in the Sec.~\ref{sec:extrapolB}, and estimates of the uncertainty are indicated by a grey band.
Experimental values are from Refs.~\cite{AjzenbergSelove:1988ec,Tilley:2002vg}.}
\label{BE2}
\end{figure}

We next examine our results for two B(E2) transitions to the ground state in Fig.~\ref{BE2}. Results as a function of  basis space parameters ($N_{\rm max}$, $\hbar\Omega$) are presented in side-by-side panels for the two interactions adopted in this work.
Similar slow convergence patterns are seen in all cases.  As may be expected,
the convergence patterns of these B(E2) transitions
are very similar to the patterns presented in Fig.~\ref{ppRMSradius} for the point proton rms radii.
As discussed above, we extrapolate the E2 matrix elements for the $(3^+,0)\rightarrow(1^+,0)$ transition and obtain the extrapolated B(E2)s
shown in the top panels of Fig.~\ref{BE2}.
The extrapolated results appear in the range of the experimental results though the extrapolation uncertainties are large.

We have not attempted to extrapolate the E2 matrix element for the $(2^+,0)\rightarrow(1^+,0)$ transition
as the $(2^+,0)$ if a broad resonance and less converged in our calculations than the $(3^+,0)$ as discussed above.
From the appearance of the results in Fig.~\ref{BE2}, it seems that
the B(E2) for the $(2^+,0)\rightarrow(1^+,0)$ transition will likely converge to a value above the experimental value.
We also note the erratic pattern in the B(E2) for the $(2^+,0)\rightarrow(1^+,0)$ transition at low $\hbar\Omega$ values for $N_{\rm max}=10$ and 12.
This is related to the significant mixing of the $(2^+,0)$ and $(2^+,1)$ states at these basis space parameters.
The close proximity of these two states is seen in Fig.~\ref{natural} where they are observed to be crossing around $N_{\rm max}=8$ at $\hbar\Omega=17.5$~MeV.
A corresponding level crossing with NNLO$_{\rm opt}$ is seen in Fig.~\ref{natural} to occur at a significantly lower $N_{\rm max}\sim 4$ which is  below the range of $N_{\rm max}$ values for the related B(E2) plotted in Fig.~\ref{BE2}.

\begin{figure}[htb]
\centering
\includegraphics[width=0.76\linewidth]{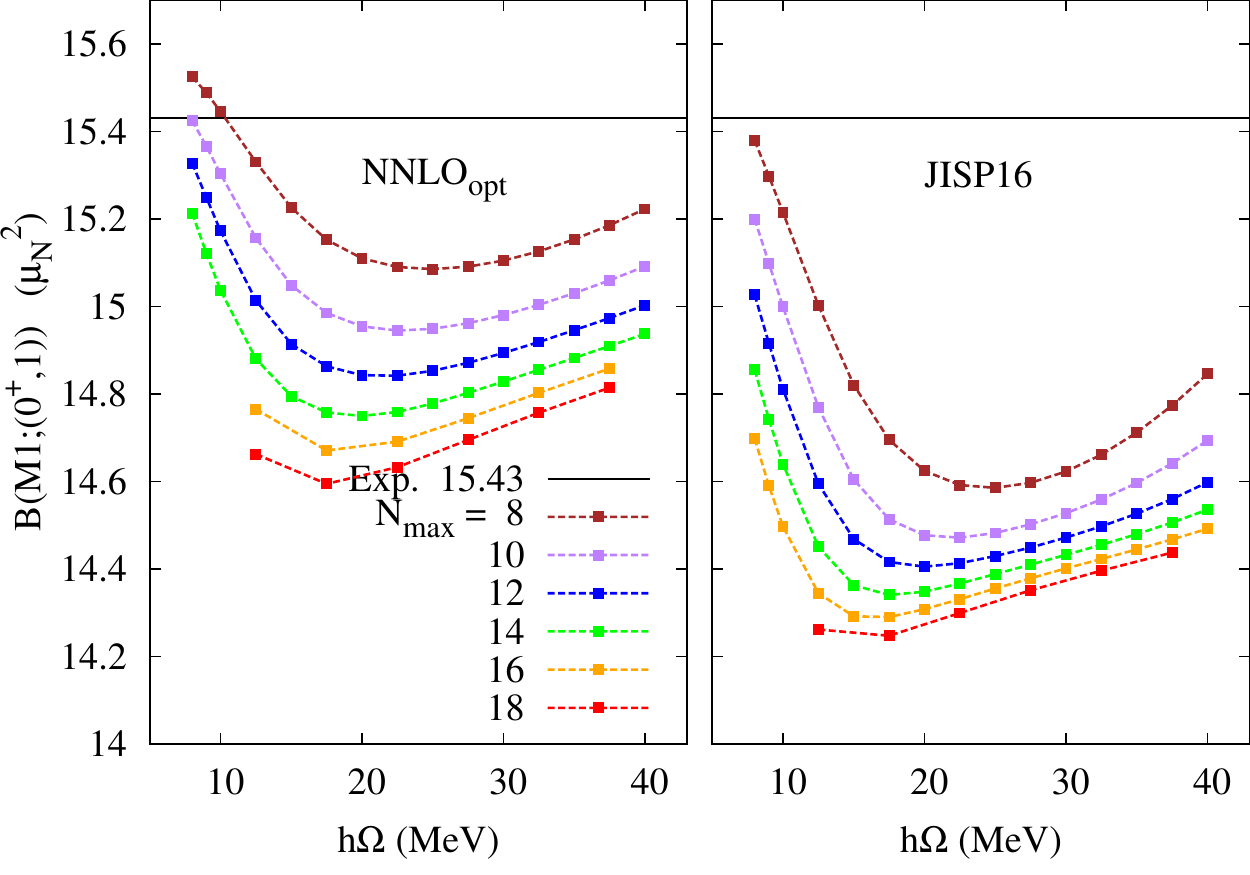}
\includegraphics[width=0.76\linewidth]{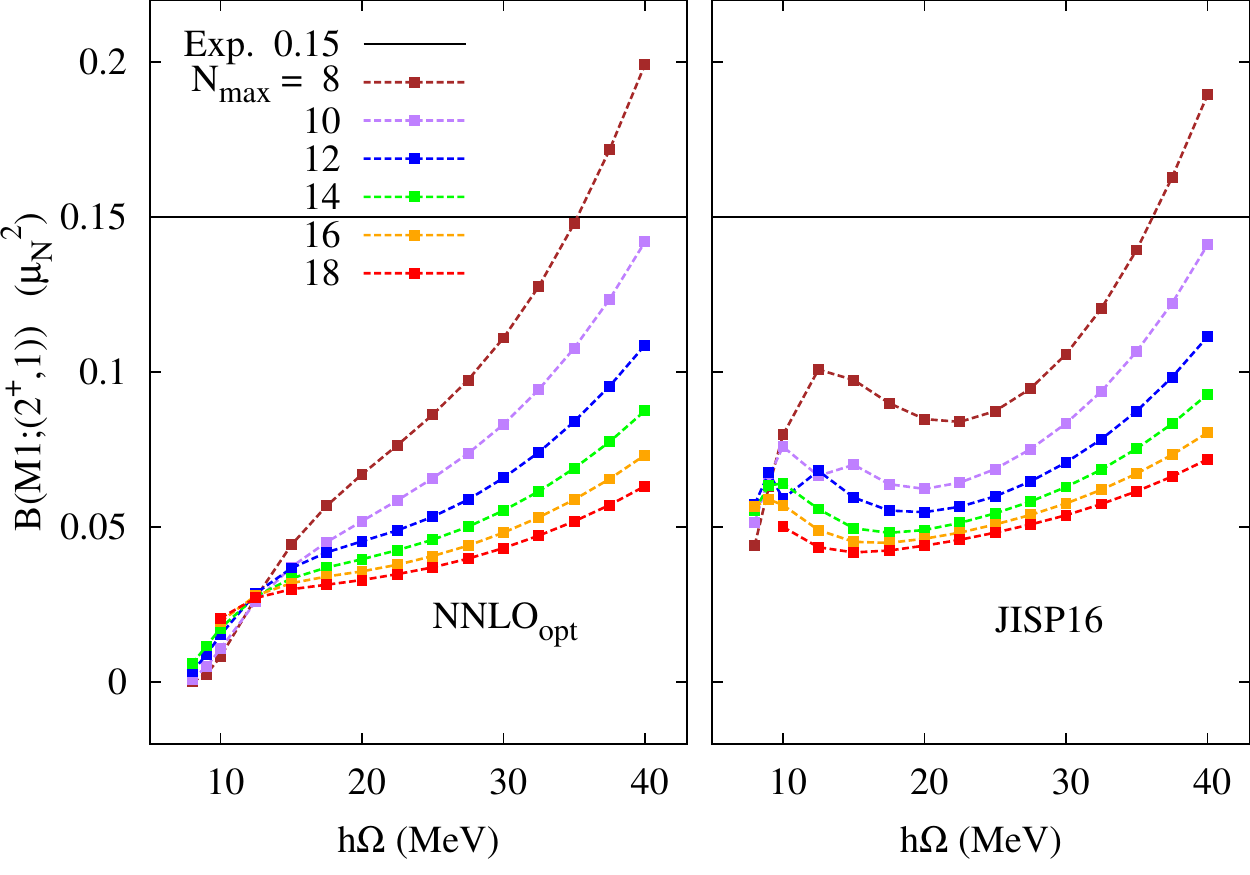}
\caption{(Color online) B(M1) to the ground state $(1^+,0)$ from $(0^+,1)$ (top) and $(2^+,1)$ (bottom) 
as a function of the HO energy for a sequence of $N_{\rm max}$ values.
Experimental values are from Refs.~\cite{AjzenbergSelove:1988ec,Tilley:2002vg}.}
\label{BM1}
\end{figure}

In Fig.~\ref{BM1} we present B(M1) transitions for $(0^+,1)\rightarrow(1^+,0)$ and $(2^+,1)\rightarrow(1^+,0)$ for both interactions as a function of the basis space parameters.
For the $(0^+,1)\rightarrow(1^+,0)$ transition the B(M1) for both interactions is trending away from experiment with increasing $N_{\rm max}$.
While the trends are systematic with increasing $N_{\rm max}$ and the changes appear small (at the 1\% level at each $N_{\rm max}$ increment), there is no apparent convergence.
For the $(2^+,1)\rightarrow(1^+,0)$ B(M1) transition in Fig.~\ref{BM1} we see reasonable convergence to a small result  compared to experiment for both interactions.
Note that the experimental result is already small compared to the
$(0^+,1)\rightarrow(1^+,0)$ B(M1) transition.
Thus, the theoretical results are approximately  consistent with the hindered nature of this transition.
One may also note the differences in the convergence patterns for the $(2^+,1)\rightarrow(1^+,0)$ transition between the two interactions used and recall the more dramatic differences seen in the quadrupole moment of Fig.~\ref{quadrupole}.

\begin{figure}[htb]
\centering
\includegraphics[width=0.9\linewidth]{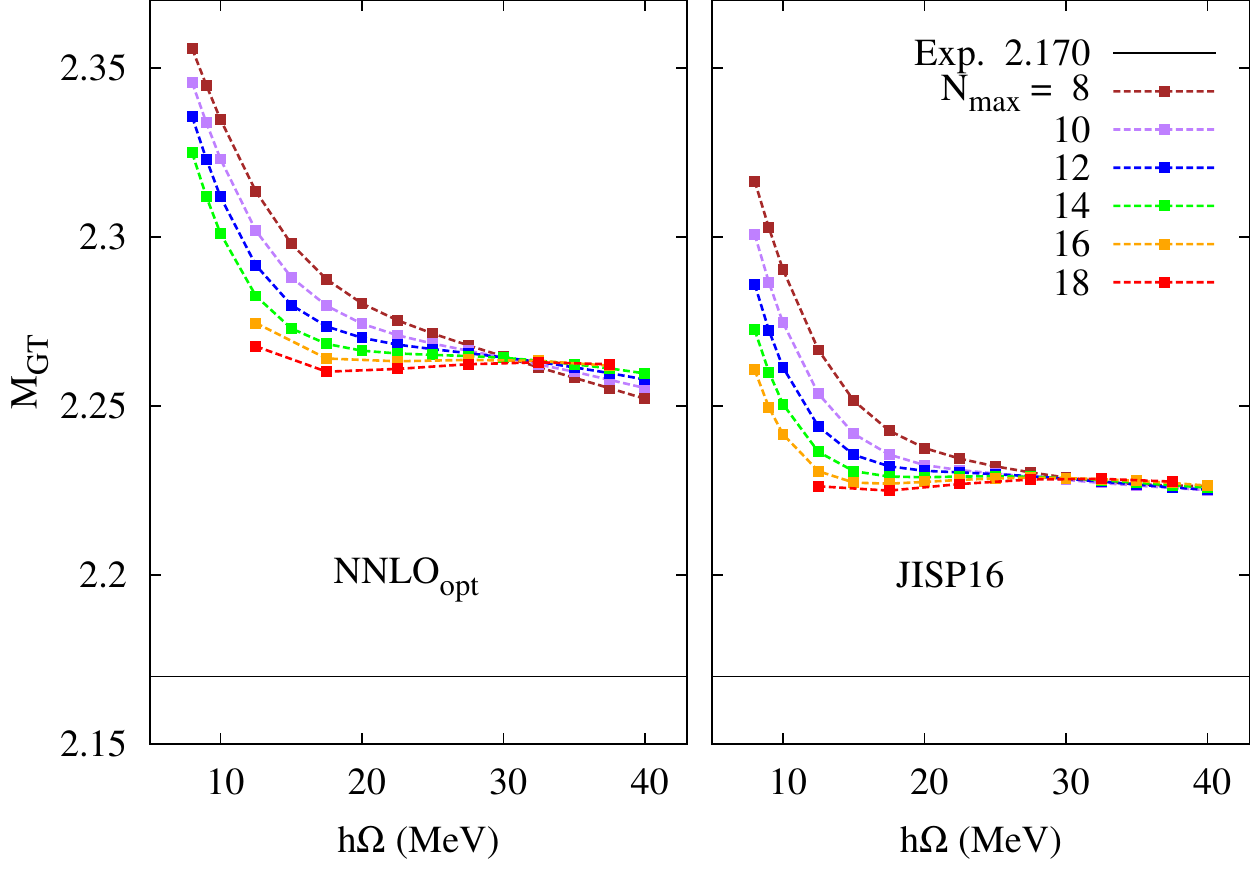}
\caption{(Color online) Gamow-Teller matrix element from $(0^+,1)$ to ground state $(1^+,0)$ 
as a function of the HO energy for a sequence of $N_{\rm max}$ values.
Note the expanded vertical scale.
The experimental value is from Ref.~\cite{Vaintraub:2009mm}.}
\label{mGT}
\end{figure}

The Gamow-Teller transition from the ground state of $^6$He to the ground state of $^6$Li has received considerable attention recently~\cite{Vaintraub:2009mm,Cockrell:2012vd,Marcucci:2008mg,Pervin:2007sc}.
We present our M$_{\rm GT}$ results for this transition using both interactions in Fig.~\ref{mGT} where we
employ the isobaric analog state $(0^+,1)$ in $^6$Li for the initial state wave function.
Both results for M$_{\rm GT}$ appear to be reasonably converged and about $3\sim 5$\% high compared with experiment.
Note that Vaintraub \textit{et al.}~\cite{Vaintraub:2009mm} have shown that chiral perturbation theory corrections to the M$_{\rm GT}$ for this transition significantly improve the agreement of the JISP16 results with experiment.

\begin{figure}[htb]
\centering
\includegraphics[width=0.73\linewidth]{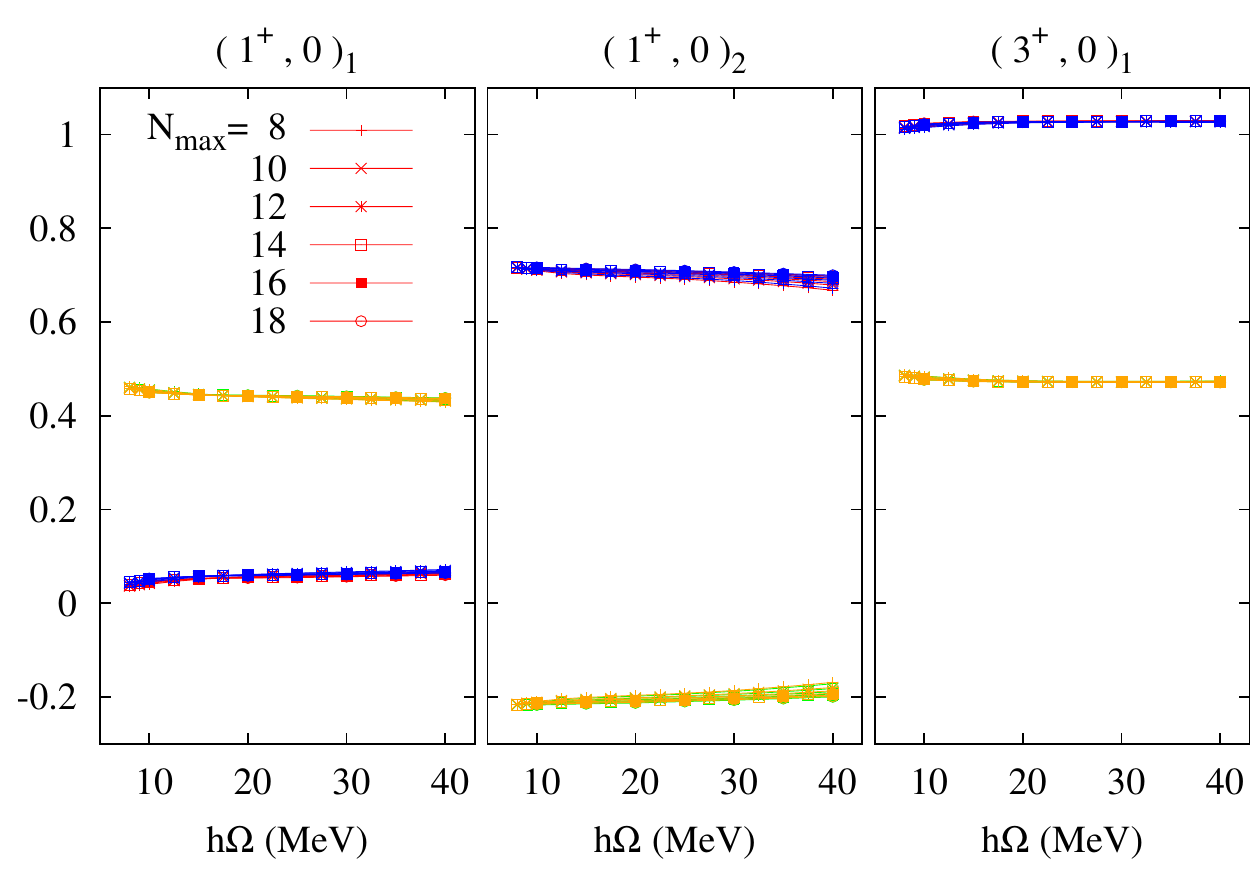}
\includegraphics[width=0.73\linewidth]{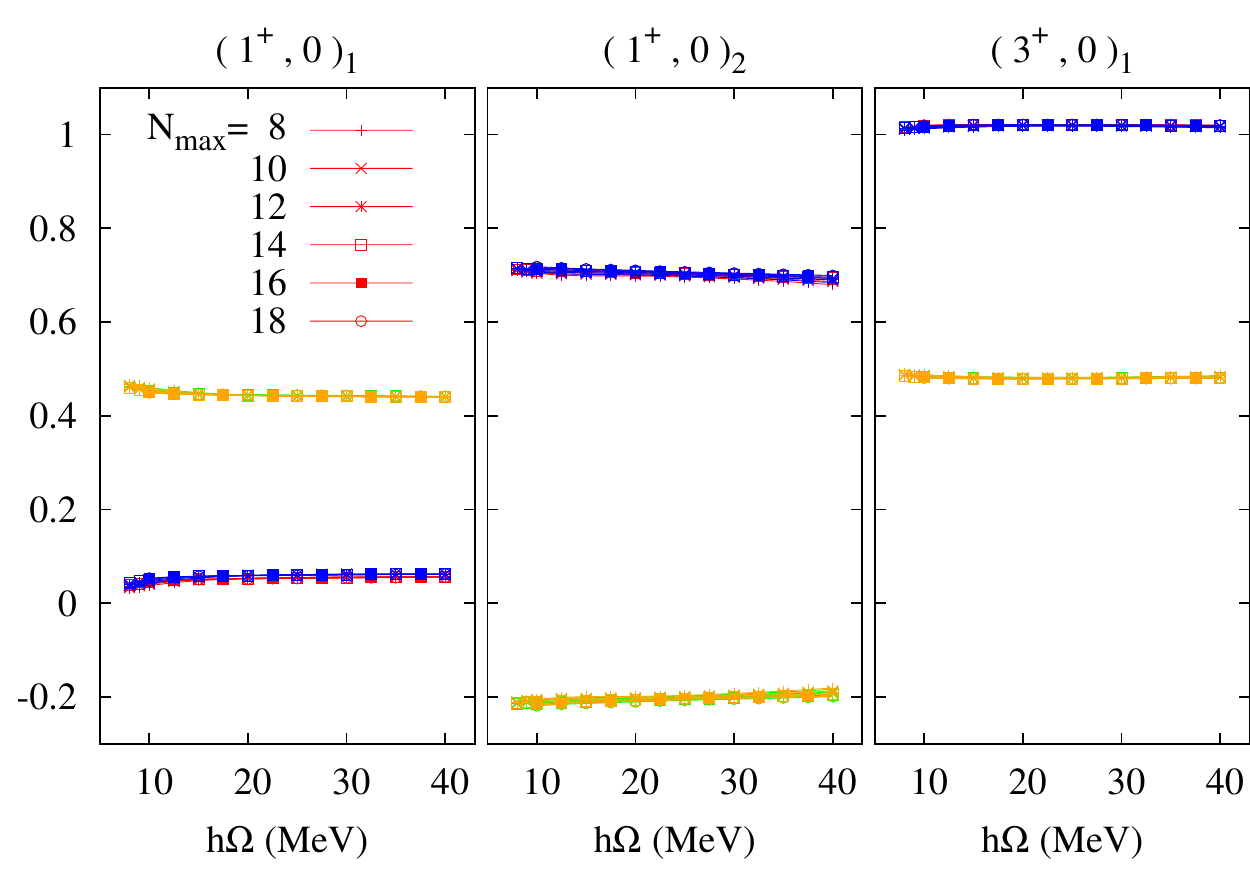}
\caption{(Color online) Spin decomposition into proton orbital motion (red), neutron orbital motion (blue), proton intrinsic spin (orange) and neutron intrinsic spin (green) of the result with NNLO$_\textrm{opt}$ (top) and JISP16 (bottom) respectively.
Selected states are ground state $1^+$, first excited $1^+$ and $3^+$ state with $T=0$.
Due to nearly exact isospin symmetry, the neutron and proton orbital motions coincide to high accuracy.
Similarly the neutron and proton spin contributions nearly coincide.}
\label{spinDecomp}
\end{figure}

In order to explore the character of a specific eigenstate, one may decompose its total angular momentum into separate orbital contributions from the protons and neutrons as well as separate spin contributions.
As explained in Ref.~\cite{Maris:2013poa} one represents the contributions to the total spin $J$ in terms of  the nucleon intrinsic spin $S$ and orbital motion $L$
\begin{equation}
 J=\frac{1}{J+1}(\langle\vec{J}\cdot\vec{L}_p\rangle+\langle\vec{J}\cdot\vec{L}_n\rangle+\langle\vec{J}\cdot\vec{S}_p\rangle+\langle\vec{J}\cdot\vec{S}_n\rangle)\, .
\end{equation}
We present in Fig.~\ref{spinDecomp} the four components of $J$ for three selected states in $^6$Li for both interactions.
Due to nearly perfect isospin symmetry, only two curves are seen in each panel.
Good convergence is also evident as there is little dependence on either $N_{\rm max}$ or $\hbar\Omega$.
The good convergence in the first panel for each interaction is linked with the good convergence of the ground state magnetic moment seen in Fig.~\ref{magneticDipole}.

By comparing these decompositions for states with the same quantum numbers,
we may identify characteristics that are sometimes useful when two states come close together
and appear to cross each other.
For example the two $(1^+,0)$ states in Fig.~\ref{spinDecomp} show very different orbital
and spin decomposition.
In addition, these characteristics can be used to see relationships between states with different angular momentum~\cite{Maris:2013poa}.
For example, the ground state $(1^+,0)_1$ and excited state $(3^+,0)$ in Fig.~\ref{spinDecomp} appear to be rotational partners as the predominant change is the  addition of one unit of orbital angular momentum each to the protons and the neutrons.  In addition, the total angular momentum of the ground state is carried equally
by the proton spin and the neutron spin indicating the likely configuration is an $\alpha$-$d$ cluster
in a relative $s$-state between the clusters in support of the conventional $\alpha$-$d$ cluster model
for the ground state of $^6$Li.

Our results for a suite of $^6$Li observables are summarized in Table~\ref{Tab6Li} and compared with
experimental results as well as with results from
JISP16 up through the Nmax = 16 model space~\cite{Cockrell:2012vd} and
AV18/IL2, the Argonne v18 NN interaction with the Illinois-2
3N interaction~\cite{Marcucci:2008mg,Pervin:2007sc}.  Experimental and theoretical uncertainties are quoted in parenthesis and represent the uncertainty in the least significant digits quoted. Overall, there is good agreement
between the {\it ab initio} theoretical results (columns 3-6) and experiment. There is also remarkable consistency among the
theoretical results with the GFMC results with AV18/IL2 showing better agreement with experiment.
This better agreement with experiment may be attributed to the role of the 3N interactions,
the GFMC method which treats long-range observables more accurately and
the incorporation of meson-exchange corrections for the magnetic observables.

With the exception of results from extrapolations discussed in Sec.~\ref{sec:extrapolB}, the NCFC results (columns 3-5) systematically underpredict the long range observables.
 Note that two different results with JISP16 (columns 4 and 5), except where we include the extrapolations from Sec.~\ref{sec:extrapolB}, agree well  with each other within their quantified uncertainties.

\begin{table}[t]
\renewcommand{\arraystretch}{1.3}
\begin{tabular}{lcrrrr}
\hline\hline
  $^6$Li                                       &~~~~~~~Exp.~~~~~~~&~~~~~~~NNLO$_{\rm opt}$ &~~~~~~~JISP16 &~~~~~~~JISP16$^*$ &~~~~~~~AV18/IL2 \\ \hline
  $E_{\rm gs}(1^+,0)$                   & 31.995     & 30.55(9)                     & 31.53(2)   &  31.49(3)      & 32.0(1) \\
  $\langle r^2_{pp}\rangle^{1/2}$ & 2.38(3)     & 2.32(9)$^\dagger$                      & 2.31(6)$^\dagger$    &   2.3            & 2.39(1)  \\[2pt]
  $E_{\rm x}(3^+,0)$                    & 2.186(2)    & 2.843(1)                     & 2.560(3)  &  2.56(2)       & 2.2(2) \\
  $E_{\rm x}(0^+,1)$                    &  3.56(1)    & 3.879(15)                    & 3.708(6)  &  3.68(6)       & 3.4(2) \\
  $E_{\rm x}(2^+,0)$                    & 4.312(22)  & 4.36(9)                       & 4.63(10)  &  4.5(3)         & 4.2(2) \\
  $E_{\rm x}(2^+,1)$                    & 5.366(15)  & 6.19(6)                       & 6.07(6)    &  5.9(2)        & 5.5(2) \\ [2pt]
  $Q(1^+,0)$                                 & -0.082(2)  & -0.034(3)$^\dagger$  & -0.072(1)$^\dagger$  & -0.077(5)    & -0.32(6) \\
  $Q(3^+,0)$                                 & -             & -5.1(3)                        & -4.8(2)    &  -4.9          &  \\ [2pt]
  $\mu(1^+,0)$                            & 0.822       & 0.8380(5)$^\ddagger$    & 0.8389(2) & 0.839(2)     & 0.800(1) \\
  $\mu(3^+,0)$                            & -             & 1.8607(1)$^\ddagger$   & 1.8654(1) &  1.866(2)    & \\ [2pt]
  B(E2;$(3^+,0)$)                          & 10.7(8)     & 12(2)$^\dagger$       & 9.2(6)$^\dagger$   & 6.1            & 11.65(13) \\
  B(E2;$(2^+,0)$)                          & 4.4(23)     &  6.6(7)$^\ddagger$       & 6.7(7)      & 7.5             & 8.66(47) \\ [2pt]
  B(M1;$(0^+,1)$)                         & 15.43(32)  & 14.59(8)                       & 14.25(4)   & 14.2(1)        & 15.02(11) \\
  B(M1;$(2^+,1)$)                         & 0.15         & 0.031(3)                       & 0.042(3)   &  0.05(1)       & \\ [2pt]
  $M_{\rm GT}$                         & 2.170        & 2.260(4)                       & 2.225(2)   & 2.227(2)      & 2.18(3)  \\
\hline\hline
\end{tabular}
\caption{
  Experimental results for $^6$Li observables and corresponding
  theoretical results using different interactions and many-body
  methods. Energies are in MeV, point proton rms radii in femtometers,
  Q in $e {\rm fm}^2$,
  $\mu$ in $\mu_N$,
  $B(E2)$ in $e^2{\rm fm}^4$ and $B(M1)$ in $\mu_N^2$.
  Columns 3 and 4 present our NCFC results based on calculations up through
  $N_{\max}=18$ with the NNLO$_{\rm opt}$ and JISP16 interactions
  respectively. We indicate with a dagger ($\dagger$) the results from extrapolations discussed in Sec.~\ref{sec:extrapolB}.
  Column 5, denoted by JISP16 with the asterisk ($*$),  quotes NCFC results based on calculations up through $N_{\max}=16$ with the JISP16 interaction and their stated
  extrapolation methods~\cite{ Cockrell:2012vd}.
  The last column presents $^6$Li observables from the Greens Function Monte Carlo (GFMC) approach using the Argonne v18 NN interaction with the Illinois-2 3N interaction (AV18/IL2)~\cite{Marcucci:2008mg,Pervin:2007sc}.  Dipole observables obtained with GFMC using AV18/IL2 also include meson-exchange corrections. Where available, all quantities have their uncertainties quoted in parenthesis for the least significant figures.
  Our ground state energies are based on extrapolation A5. All other observables in columns 3 and 4, not flagged with a ``dagger'' or a ``double dagger'', are calculated at $N_{\rm max}=18$
  and $\hbar\Omega=17.5$~MeV with uncertainties defined as differences between
  $N_{\rm max}=16$ and 18.
  However results indicated with a double dagger ($\ddagger$) are calculated at $\hbar\Omega=20$~MeV.
  All transitions are to the ground state $(1^+,0)$.
  The references for experimental values are~\cite{Tanihata:2013jwa} for rms radius,~\cite{Vaintraub:2009mm} for Gamow-Teller matrix element, and~\cite{AjzenbergSelove:1988ec,Tilley:2002vg} for others.
}
\label{Tab6Li}
\end{table}

\section{Ab initio Gamow Shell Model}
As one approaches the particle emission thresholds, it becomes
increasingly important to take into account the coupling to the decay and scattering channels. The recently developed
complex-energy Gamow Shell Model (GSM)~\cite{gsm_rev} has proven to be a
reliable tool in the description of nuclei where these continuum effects
cannot be neglected. In the GSM, the coupling to the continuum is taken into account by working with a many-body basis 
 constructed from a s.p.~Berggren ensemble~\cite{Berggren:1968zz}  which includes bound,
resonant and continuum states. In practice, 
one discretizes the continuum states  to obtain a finite set of s.p.~shells and, as in any Shell Model
calculation, the dimension of the Hamiltonian matrix grows rapidly with
the number of s.p.~states and  nucleons. Moreover, the GSM Hamiltonian matrix being non-Hermitian, 
advanced numerical methods that can handle large non-Hermitian matrices are needed in this context.
It has been shown that the Density Matrix Renormalization Group (DMRG)
offers an efficient way to solve this problem at a relatively low computational cost~\cite{dmrg}.

Let us go into more details on the application of DMRG in the $J$-scheme in the context
of the GSM. The objective is to calculate an eigenstate
$|J^{\pi}\rangle$ of the GSM Hamiltonian $\hat{H}$ with angular
momentum $J$ and parity $\pi$. The s.p.~states are usually taken as solutions of the Hartree-Fock (HF) potential for the lowest angular
momentum $l$ whereas for higher values of $l$, the HO shells are considered.
As $|J^{\pi}\rangle$ is a many-body
pole of the scattering matrix of $\hat{H}$, the contribution from
non-resonant scattering shells along the contours 
are usually smaller than the contribution
from the s.p.~poles {\it i.e.} the bound and resonant shells~\cite{gsm_rev}. Based on this observation, one defines a subspace $A$ (the ``reference subspace")
which contains states constructed from the poles and a complementary subspace $B$, which contains states constructed from the other
shells~\cite{dmrg}. At each DMRG iteration, one will construct and optimize the subspace $B$ while $A$ is kept fixed.

First, one constructs the states $|k\rangle_A$ which form the reference
subspace $A$.  All possible matrix elements of operators of the GSM
Hamiltonian $\hat{H}$ acting in $A$ are calculated and stored. The GSM
Hamiltonian is  then diagonalized in $A$ to provide a
zeroth-order approximation $|\Psi_J\rangle^{(0)}$ to
$|J^{\pi}\rangle$. 
The rest of the shells are then gradually added one by one during the first stage of the DMRG procedure, the so-called warm-up phase.
Let us assume we have reached the $n^{th}$ iteration where the shell $n$ is added. 
At that point, one constructs all possible many-body states $|i\rangle_B$ in $B$ by coupling previously optimized states in $B$ 
and states constructed by occupying the shell $n$. All matrix elements of operators of the GSM Hamiltonian acting on
these new states in $B$ are also computed.
By coupling the states  $|k\rangle_A$  with the states $|i\rangle_B$, one
constructs the set of states of a given $J^{\pi}$. Recent work to
perform accurate and fast evaluation of spin-coupling
coefficients~\cite{Johansson:2016} has allowed an important speedup of this step.
This ensemble
serves as a basis in which the GSM Hamiltonian is diagonalized.  The
target state $|\Psi_J\rangle$  is selected among the eigenstates of
$\hat{H}$ as the one having the largest overlap with the reference
vector $|\Psi_J\rangle^{(0)}$. Then, a truncation is
performed in $B$ by introducing the reduced density matrix $\rho^B$,
constructed by summing over the reference subspace
$A$. The reduced density matrix being complex-symmetric,
the truncation is done by keeping the eigenstates
$\alpha_{B} $ (the `optimized' states) with the largest moduli
of eigenvalue $w_{\alpha}$~\cite{dmrg}. More precisely, we keep as
many eigenstates of $\rho^B$ such that the condition $|1-\sum_{\alpha}w_{\alpha}|\leq \epsilon$ is satisfied. The quantity $\epsilon$ here can be viewed
as the truncation error of the reduced density matrix.

As the warm-up phase ends, the so-called sweeping phase begins.  Starting from the last added shell, 
the procedure continues in the reverse direction (the `sweep-down'
phase) until the first scattering shell is reached. The procedure is
then reversed and a sweep in the upward direction (the `sweep-up'
phase) begins. The sweeping sequences continue until convergence for
the target eigenvalue is achieved.

We show now an application of the DMRG method for the description of the bound $J^{\pi}=1^+$  ground state 
and  the resonant  $J^{\pi}=2_2^+$ state  in $^6\rm{Li}$.
The $J^{\pi}=2_2^+$ resonance being above the thresholds for neutron and proton emission, we include in the s.p.~basis, the $s_{1/2}$ 
bound state as well as the $p_{3/2}$ resonance given by the HF potential 
(solved in the spherical approximation) with the JISP16 interaction, for both proton and neutron. 
For other partial waves, the HF potential either has resonant pole at high energy whose contribution to the many-body states is negligible or
does not have resonant states. In these cases, the s.p.~bases are taken as the  HO shells.
 In this application, we include  HO shells up to the $g$ shells ($l=4$) and with a total  energy equal to $10 \hbar \omega$.
The s.p.~bound states for the $s_{1/2}$ neutron and proton are respectively at  $-27.406$~MeV and $-25.819$~MeV. 
The position of the $p_{3/2}$ neutron resonance is $(0.180,-0.027)$~MeV whereas the $p_{3/2}$ proton resonance is located at 
$(1.626,-0.341)$~MeV. In order to fulfill the Berggren completeness relation, the  $s_{1/2}$ real-continuum scattering 
states and $p_{3/2}$  complex-scattering states are included. These continua are discretized with 15 and 25 points for 
the $s_{1/2}$  and $p_{3/2}$ contours respectively. Although a pure HO basis would be sufficient to describe the bound $J^{\pi}=1^+$ ground state, we have
 chosen to use the same s.p.~basis for both  the ground state and the resonance in order to calculate these states consistently in the same many-body space.
In order to mitigate the computational cost of the DMRG iterations, an additional truncation is introduced here. As we wrote previously, the many-body states in $B$
are generated from a s.p.~set containing HF and HO shells. We will restrict the subspace $B$
to states that have a component generated by the HO shells of the s.p.~basis with an average HO energy (in modulus) less than or equal to $10\hbar\omega$. This criteria will
also allow us to compare our results with the NCSM calculations in the HO basis.

\begin{figure}[htb]
\begin{center}
\includegraphics[scale=0.6]{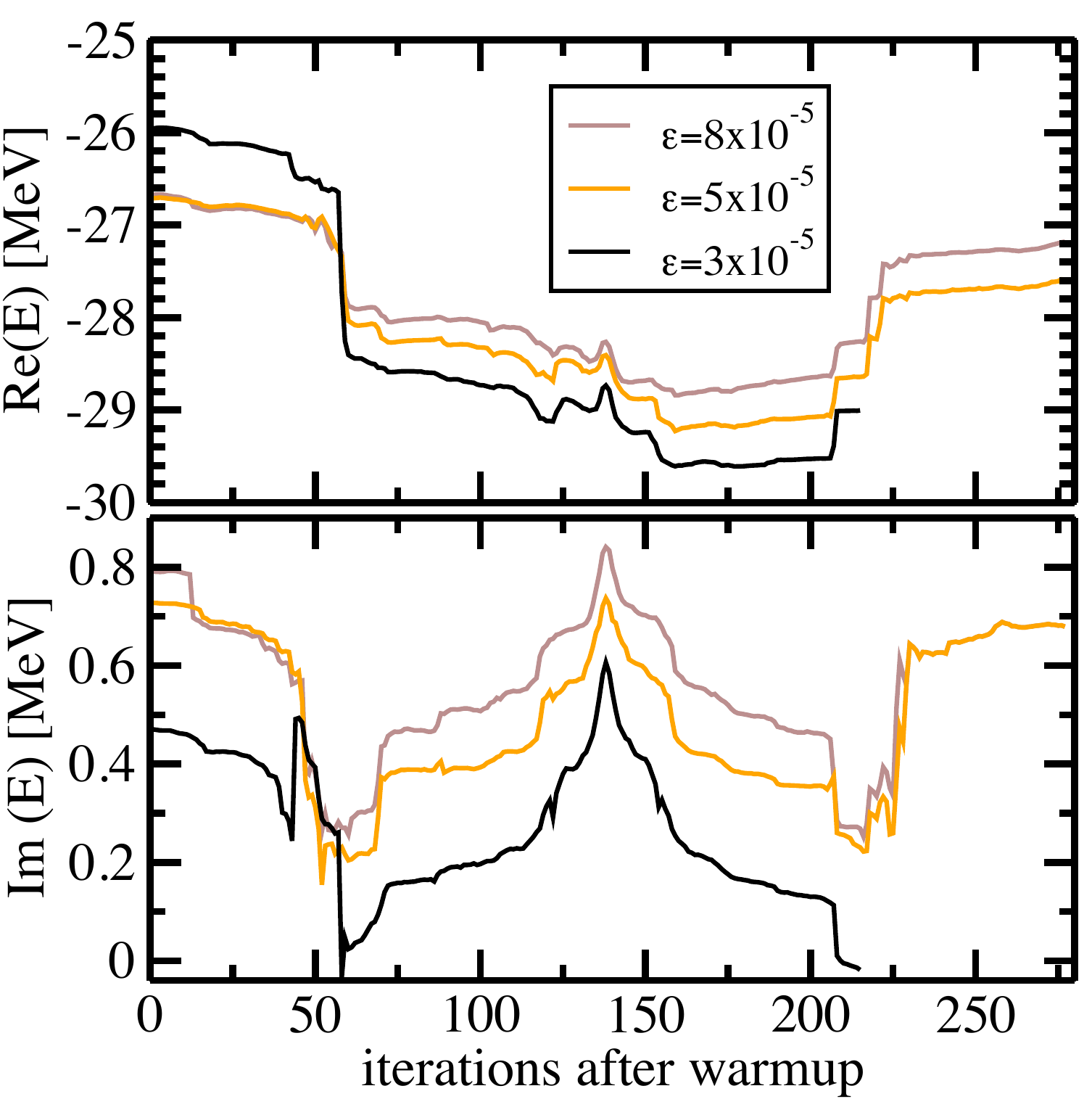}
\caption{Real and imaginary parts of the JISP16 ground state energy in $^6{\rm{Li}}$ given by  the DMRG approach
for $\epsilon = 8 \times 10^{-5}$, $5 \times 10^{-5}$ and $3 \times 10^{-5}$. Partial waves up to $l = 4$
(s,p,d,f,g) are included. 
 Results are shown starting from the first sweep until the end of the second sweep for $\epsilon = 8 \times 10^{-5}$ and $5 \times 10^{-5}$ 
whereas for  $\epsilon =3 \times 10^{-5}$  results are only partially shown during the second sweep.}
\label{fig:dmrg_Li6_re_im}
\end{center}
\end{figure}

In Fig.~\ref{fig:dmrg_Li6_re_im} we show results for the  $J^{\pi}=1^+$ ground state with truncations given by 
 $\epsilon= 8 \times 10^{-5}$, $5 \times 10^{-5}$ and $3 \times 10^{-5}$. Results are shown starting at the first sweep until the end of the second sweep excepting for $\epsilon=3 \times 10^{-5}$ where results are only partially shown during the second sweep. Clearly, the energy does not converge as the number of iterations increases. Moreover the 
imaginary part of the energy, which is expected to be zero, does not vanish. For
 $\epsilon=5 \times 10^{-5}$, the real part of the energy varies from $-28.406$~MeV to $-27.590$~MeV during the second sweep which corresponds 
to the iterations $\#138$ to $\#277$. In that region the imaginary part fluctuates between $0.730$~MeV and  $0.680$~MeV.

In order to improve convergence we have considered, for the first time here, a new approach based on 
using an optimized s.p.~basis. For a fixed $\epsilon$, we first perform a DMRG calculation up to the end of the warm-up. At that point, we
obtain an approximation of the many-body state $|\Psi_J\rangle$ from which we calculate the one-body density 
matrix $\rho^{\rm 1-b}_{\alpha,\beta}=\langle \Psi_J| a^{\dagger}_{\alpha}\tilde{a}_{\beta}|\Psi_J\rangle$. We then diagonalize $\rho^{\rm 1-b}$ and obtain the so-called natural orbitals which will form a new set of s.p.~states. We then perform DMRG calculations up to the end of the warm-up with this new set and again diagonalize the one-body density  matrix. Once convergence for the s.p.~basis has been achieved, we continue the calculations for the sweeeping phase.
 It should be noted that similar optimization of the s.p.~basis based on diagonalization of the one-body density matrix
or the resolution of  the Generalized  Brillouin equation~\cite{bri} have been reported for nuclei in the context of the 
Variational Multiparticle-Multihole Configuration Mixing Method~\cite{pillet}.

\begin{figure}[htb]
\begin{center}
\includegraphics[scale=0.6]{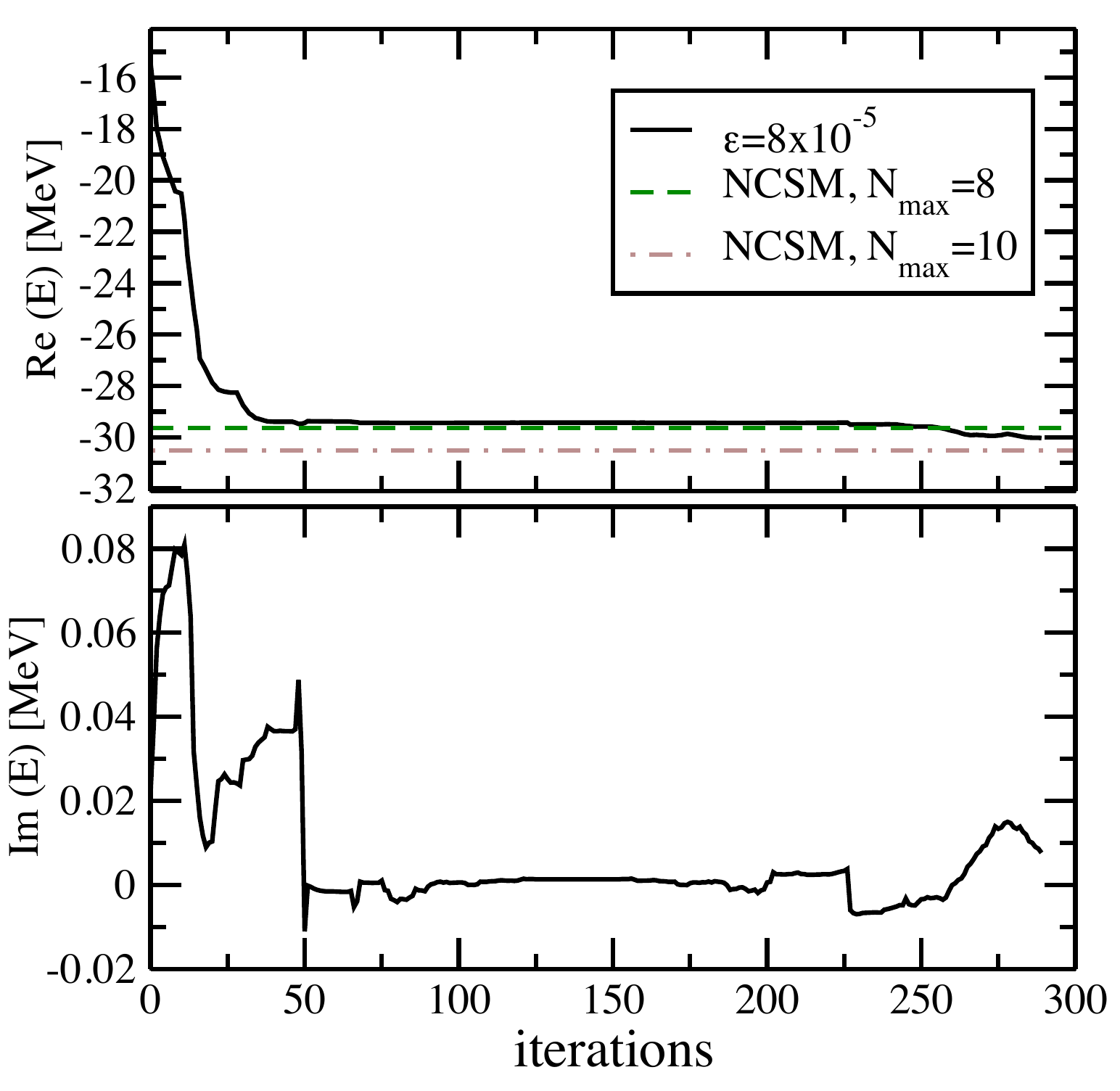}
\caption{Real and imaginary parts of the JISP16 ground state energy given by the DMRG approach
for $\epsilon = 8 \times 10^{-5}$ using a set of optimized shells. 
Results are shown starting from the warm-up  until the beginning of the second sweeping phase.
For comparison results obtained with the NCSM with HO shells are shown for $N_{\rm max}=8$ and $10$.}
\label{fig:dmrg_Li6_8_e5_opt}
\end{center}
\end{figure}

Results for $\epsilon=8 \times 10^{-5}$, after having performed three warm-up calculations to optimize the s.p.~basis, 
 are shown in Fig.~\ref{fig:dmrg_Li6_8_e5_opt} for the real and imaginary part of the ground state energy. 
 Results are shown starting from the warm-up phase until the beginning of the second sweep. Clearly, the convergence is greatly improved 
by using a set of optimized s.p.~states. The lowest values for the
energy is equal to $-30.027$~MeV which is reached at  the last iteration ($\#289$) in Fig.~\ref{fig:dmrg_Li6_8_e5_opt}.
The imaginary part is drastically reduced and  varies from $1$~keV to $14$~keV during the sweeping phase.
Although the energy could certainly be improved by pursuing the calculation even further during the second sweeping phase, we can nevertheless assume,
based on our previous experience of applications of DMRG in the context of the GSM~\cite{dmrg}, that the energy obtained at the last iteration in Fig.~\ref{fig:dmrg_Li6_8_e5_opt} gives
a good estimate of the total energy of the ground state in $^6{\rm Li}$. Moreover we expect that the changes of the wave function and the total energy during the sweeping phases to be smaller when natural orbitals are considered. This can be seen for instance by comparing the fluctuations for $\epsilon = 8 \times 10^{-5}$ between the beginning and the end of the sweeping
down phase using non-optimized orbitals (see Fig.~\ref{fig:dmrg_Li6_re_im}) and optimized orbitals (see Fig.~\ref{fig:dmrg_Li6_8_e5_opt}):
 in the first case, the real part of the energy varies from $-26.675$~MeV to $-28.314$~MeV
 whereas for calculations with optimized orbitals, the real part of the energy varies from $-29.432$~MeV to $-29.866$~MeV. This can be understood by the fact that a portion of the many-body correlations
are already effectively taken into account at the s.p.~level by using  natural orbitals.

We show in Fig.~\ref{fig:dmrg_Li6_opt} results using the set of optimized shells for $\epsilon=5 \times 10^{-5}$,
$2 \times 10^{-5}$ and $1 \times 10^{-5}$ during the warm-up phase. The lowest energy obtained for $\epsilon=1 \times 10^{-5}$ 
is equal to $-30.150$~MeV at iteration \#155  whereas the imaginary
part (not shown) at that iteration is equal to $-13$~keV. For 
comparison we show also the results obtained with the NCSM at $N_{\rm max}=8$ and $N_{\rm max}=10$. We can see  
that for the largest calculations considered here {\it i.e.}  
$\epsilon = 2 \times 10^{-5}$ and $1 \times 10^{-5}$, the lowest energies are below the NCSM result at $N_{\rm max}=8$ and above the NCSM energy at $N_{\rm max}=10$.
In the NCSM, truncations at a given $N_{\rm max}$ are defined with respect to the lowest configuration. For $^6{\rm{Li}}$, the lowest
configuration has $N=2 \hbar \omega$ (4 nucleons in the $0s_{1/2}$ shells and one proton and one neutron in  $0p_{3/2}$). As a consequence, 
for the NCSM calculations at $N_{\rm max}$, all basis states with an HO energy less {than or equal to $(N_{\rm max}+2)\hbar \omega$  are included.
Due to the additional truncation on the states in $B$ based on their components in the HO basis
 (states with components on the HO shells belonging to the s.p.~basis that have an energy larger than $10 \hbar \omega$ are not included in $B$),
it is not surprising that the DMRG result has an energy 
above the NCSM at $N_{\rm max}=10$. 
On the other hand, the DMRG result is below the NCSM at $N_{\rm max}=8$. This can be understood by the fact that during the DMRG iterations,
some contribution by states with energy higher than $10 \hbar \omega$ are effectively included in $B$. For instance, states with 
a non-zero occupation number of the HF shells will have a component on HO states with energies larger than $10 \hbar \omega$.

\begin{figure}[htb]
\begin{center}
\includegraphics[scale=0.4]{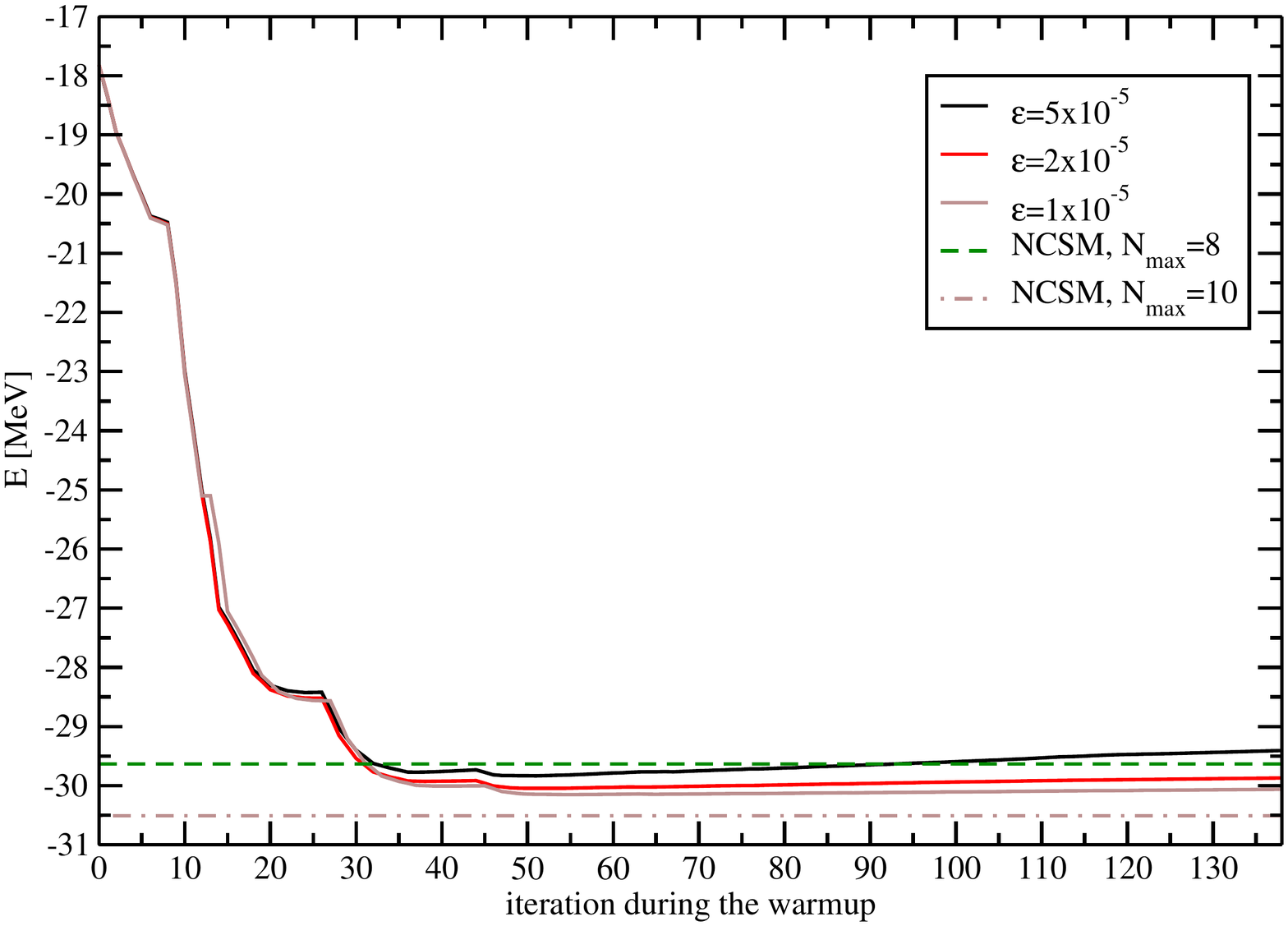}
\caption{Real  part of the JISP16 ground state energy given by the DMRG approach
for $\epsilon = 5 \times 10^{-5}$, $2 \times 10^{-5}$ and $1 \times 10^{-5}$ using a set of optimized shells. 
Results are shown for iterations during the warm-up phase.
For comparison results obtained with the NCSM with HO shells are shown for $N_{\rm max}=8$ and $10$.}
\label{fig:dmrg_Li6_opt}
\end{center}
\end{figure}

We now show in Fig.~\ref{fig:dmrg_Li6_J2_opt} the real and imaginary
part of the $J^{\pi}=2_2^+$ resonance in $^6\rm{Li}$ during the
warm-up phase for $\epsilon= 8 \times 10^{-5}$, $7 \times 10^{-5}$,
$4 \times 10^{-5}$ and $3 \times 10^{-5}$ using the set of optimized
shells.  Following the complex variational principle~\cite{dmrg}, we
select the best estimate for the energy at the DMRG iteration where
the modulus square of the energy is optimal. In our case, this
corresponds to $E=(-24.010,0.019)$~MeV reached at iteration \#52 for
$\epsilon= 3 \times 10^{-5}$. As for the ground state, the real part
of the energy is below the NCSM result at $N_{\rm max}=8$ and
(slighlty) above the NCSM result at $N_{\rm max}=10$. 
The positive imaginary part of the energy corresponds to a decay width
equal to $-38$~keV which is clearly unphysical whereas the experimental width of the state is $541$~keV. 
This negative value for the calculated width is most likely due to the discretization of the contours 
where a larger number of points would seem to be preferred.
The discretization effect also appears for
the ground state where one should expect a vanishing imaginary part
whereas the best results obtained with DMRG give a positive width of
$26$~keV.

\begin{figure}[htb]
\begin{center}
\includegraphics[scale=0.6]{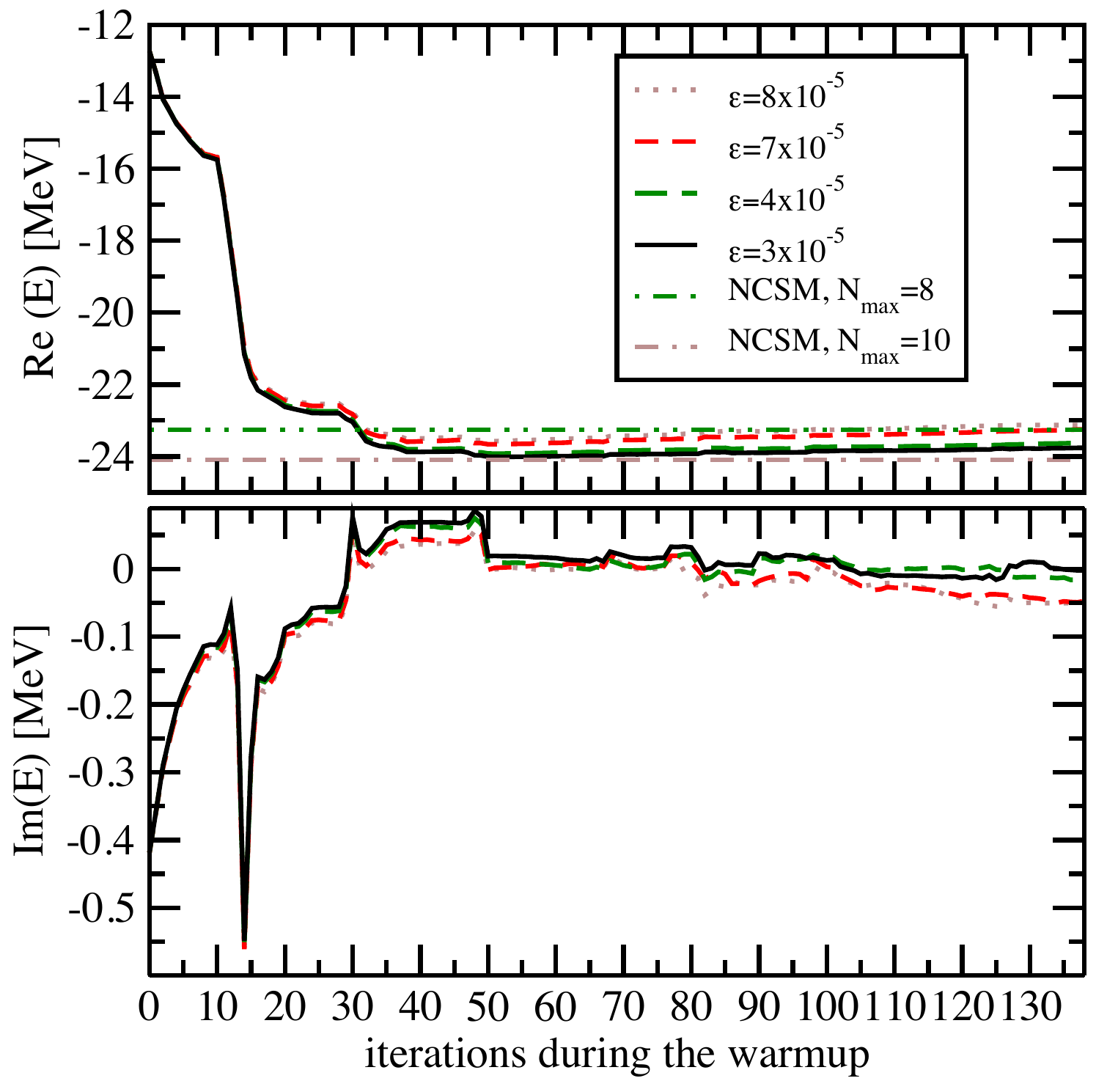}
\caption{Real and imaginary parts of the JISP16 $J^{\pi}=2_{2}^{+}$ resonance energy in $^6\rm{Li}$ given during the warm-up phase
for $\epsilon= 8 \times 10^{-5}$, $7 \times 10^{-5}$, $4 \times 10^{-5}$ and $3 \times 10^{-5}$ using a set of optimized shells.
 For comparison results obtained with the NCSM with HO shells are shown for $N_{\rm max}=8$ and $10$.}
\label{fig:dmrg_Li6_J2_opt}
\end{center}
\end{figure}

\section{Summary}

We have calculated the properties of $^6$Li using the {\it ab initio}
No-Core Full Configuration (NCFC) method with the JISP16 and
NNLO$_{\rm opt}$ potentials in which finite matrix results are
calculated up through $N_{\rm max}=18$.  We solved for both natural
and unnatural parity states.  We obtained the ground state energies,
excitation energies, point proton rms radii, magnetic and quadrupole
moments, E2 and M1 transition rates, and Gamow-Teller matrix elements.
We also improved the extrapolation methods to achieve a good
estimation of our ground state energies, point proton rms radii,
ground-state quadrupole moment, and the B(E2) from the first excited
$3^+$ to the ground state in the infinite matrix limit.  We introduced
methods for quantifying uncertainties and tested their consistency
with sequences of results in $^6$Li.

Our results together with experimental values and those from AV18/IL2 are compiled in Table~\ref{Tab6Li}.
For many observables our results are sufficiently converged to reveal where omitted 3N interactions are
expected to make significant contributions that will help close the gap between these NCFC results and experiment.
Overall, our NCFC results compare favorably with experiment
and with results from AV18/IL2 after taking into account our quantified uncertainties and our neglect of 3N interactions.

In order to investigate further the character of the ground state and two low-lying solutions, we decomposed the total angular momentum of these states into separate orbital contributions of the protons and neutrons as well as separate spin contributions.  
This decomposition supports an $\alpha$-$d$ cluster structure picture for the $(1^+,0)$ ground
state and excited $(3^+,0)$.  Furthermore, the excited $(3^+,0)$ carries the increase in orbital angular momentum that is a signature for the rotational excitation of the $(1^+,0)$ ground state.

We then investigated the ground state and $J^{\pi}=2_{2}^{+}$ resonance of $^6$Li within the Gamow Shell Model (GSM) plus the Density Matrix Renormalization Group (DMRG) approach using the JISP16 potential to establish a baseline case for future investigations of $^6$Li resonant states.
In this work, we found the conventional methods of iteratively developing the GSM+DMRG basis did now work as well as we had hoped. 
We then found that the natural orbital s.p.~basis approach succeeded in producing a well-converged ground state energy in the Berggren basis.
Similar qualitative behaviour was shown for the real part of the energy of the $J^{\pi}=2_{2}^{+}$ resonance
whereas the decay width obtained in that case was slightly negative.
We suspect that this unphysical result for the width is due to an insufficient number of discretized continuum states included in the s.p.~basis. 
Nevertheless, we think that our findings using natural orbital s.p.~basis opens a promising pathway for further applications to resonant states in light nuclei.

\section*{Acknowledgments}
We acknowledge beneficial discussions with Andrey Shirokov, Mark
Caprio, and Thomas Papenbrock, and we thank H.~T.~Johansson for
assistance with implementing fast and accurate computation of Wigner
$9j$ symbols. We also acknowledge George Papadimitriou for helpful
discussions and for help in detecting/correcting an error in one of
the codes.
The work of YK and IJS was supported  by the Rare Isotope Science Project of Institute for Basic Science funded by Ministry of Science, ICT and Future Planning and National Research Foundation of Korea (2013M7A1A1075764).
This work was supported in part by  
the U.S. Department of Energy under Grant
Nos. DE-FG02-87ER40371 and DESC0008485 (SciDAC/NUCLEI).
The research leading to these results has received funding from the
European Research Council under the European Community's Seventh
Framework Programme (FP7/2007-2013) / ERC grant agreement no.~240603.
A portion of the computational resources were provided by the National
Energy Research Scientific Computing Center (NERSC), which is supported
by the U.S. DOE Office of Science under Contract
No. DE-AC02-05CH11231.
Some computations were performed with resources provided by the Swedish
National Infrastructure for Computing (SNIC) at HPC2N.
Computational supercomputing resources were also provided by the Supercomputing Center/Korea Institute of Science and Technology Information
including technical support (KSC-2012-C3-054).


\begin{thebibliography}{99}

\bibitem{Maris:2008ax}
  P.~Maris, J.~P.~Vary and A.~M.~Shirokov,
  Phys.\ Rev.\ C {\bf 79}, 014308 (2009),
  and references therein.

\bibitem{Shirokov:2003kk}
  A.~M.~Shirokov, A.~I.~Mazur, S.~A.~Zaytsev, J.~P.~Vary and T.~A.~Weber,
  Phys.\ Rev.\ C {\bf 70}, 044005 (2004).

\bibitem{Shirokov:2004ff}
  A.~M.~Shirokov, J.~P.~Vary, A.~I.~Mazur, S.~A.~Zaytsev and T.~A.~Weber,
  Phys.\ Lett.\ B {\bf 621}, 96 (2005).

\bibitem{Shirokov:2005bk}
  A.~M.~Shirokov, J.~P.~Vary, A.~I.~Mazur and T.~A.~Weber,
  Phys.\ Lett.\ B {\bf 644}, 33 (2007).

\bibitem{Ekstrom:2013kea}
  A.~Ekstr\"om, G.~Baardsen, C.~Forss\'en, G.~Hagen, M.~Hjorth-Jensen, G.~R.~Jansen, R.~Machleidt and W.~Nazarewicz {\it et al.},
  Phys.\ Rev.\ Lett.\  {\bf 110}, 192502 (2013).

\bibitem{Cockrell:2012vd}
  C.~Cockrell, J.~P.~Vary and P.~Maris,
  Phys.\ Rev.\ C {\bf 86}, 034325 (2012).

\bibitem{Maris:2013poa}
  P.~Maris and J.~P.~Vary,
  Int.\ J.\ Mod.\ Phys.\ E {\bf 22}, 1330016 (2013).

\bibitem{Shirokov:2014rvw}
  A.~M.~Shirokov, V.~A.~Kulikov, P.~Maris and J.~P.~Vary,
  in Nucleon-Nucleon and Three-Nucleon Interactions,
  edited by L.D. Blokhintsev and I.I. Strakovsky, Nova Science, Chapter 8, p.231 (2014).

\bibitem{Barrett:2013nh}
  B.~R.~Barrett, P.~Navr\`atil and J.~P.~Vary,
  Prog.\ Part.\ Nucl.\ Phys.\  {\bf 69}, 131 (2013).

\bibitem{gsm_rev}
  N.~Michel, W.~Nazarewicz, M.~Ploszajczak and K.~Bennaceur,
  Phys.\ Rev.\ Lett.\  {\bf 89}, 042502 (2002);
  N.~Michel, W.~Nazarewicz, M.~Ploszajczak and J.~Okolowicz,
  Phys.\ Rev.\ C {\bf 67}, 054311 (2003);
  N.~Michel, W.~Nazarewicz, M.~Ploszajczak and T.~Vertse,
  J.\ Phys.\ G {\bf 36}, 013101 (2009);
  G.~Papadimitriou, A.~T.~Kruppa, N.~Michel, W.~Nazarewicz, M.~Ploszajczak and J.~Rotureau,
  Phys.\ Rev.\ C {\bf 84}, 051304 (2011).

\bibitem{Berggren:1968zz} 
  T.~Berggren,
  Nucl.\ Phys.\ A {\bf 109}, 265 (1968).

\bibitem{dmrg}
  J.~Rotureau, N.~Michel, W.~Nazarewicz, M.~Ploszajczak and J.~Dukelsky,
  Phys.\ Rev.\ Lett.\  {\bf 97}, 110603 (2006);
  J.~Rotureau, N.~Michel, W.~Nazarewicz, M.~Ploszajczak and J.~Dukelsky,
  Phys.\ Rev.\ C {\bf 79}, 014304 (2009);
  G.~Papadimitriou, J.~Rotureau, N.~Michel, M.~Ploszajczak and B.~R.~Barrett,
  Phys.\ Rev.\ C {\bf 88}, no. 4, 044318 (2013).

\bibitem{Machleidt:2011zz}
  R.~Machleidt and D.~R.~Entem,
  Phys.\ Rept.\  {\bf 503}, 1 (2011).

\bibitem{Sternberg:2008:ACI:1413370.1413386}
  P.~Sternberg, E.~G. Ng, C.~Yang, P.~Maris, J.~P. Vary, M.~Sosonkina and H.~V. Le,
  {\em Accelerating configuration interaction calculations for nuclear structure},
  in {\em Proc. of the 2008 ACM/IEEE conf. on Supercomputing\/}, IEEE Press, Piscataway, NJ, pp.15:1-15:12 (2008).

\bibitem{DBLP:journals/procedia/MarisSVNY10}
  P.~Maris, M.~Sosonkina, J.~P. Vary, E.~G. Ng and C.~Yang, Proc.~Comput.~Sci.~{\bf 1}, 97 (2010).

\bibitem{CPE:CPE3129}
  H.~M. Aktulga, C.~Yang, E.~G. Ng, P.~Maris and J.~P. Vary,
  {\em Improving the scalability of symmetric iterative Eigensolver for multi-core platforms},
  Concurrency Computat.: Pract. Exper. DOI: 10.1002/cpe.3129 (2013, in press).

\bibitem{Forssen:2008qp}
  C.~Forss\'en, J.~P.~Vary, E.~Caurier and P.~Navr\`atil,
  Phys.\ Rev.\ C {\bf 77}, 024301 (2008).

\bibitem{Coon:2012ab}
  S.~A.~Coon, M.~I.~Avetian, M.~K.~G.~Kruse, U.~van Kolck, P.~Maris and J.~P.~Vary,
  Phys.\ Rev.\ C {\bf 86}, 054002 (2012).

\bibitem{Furnstahl:2012qg}
  R.~J.~Furnstahl, G.~Hagen and T.~Papenbrock,
  Phys.\ Rev.\ C {\bf 86}, 031301 (2012).

\bibitem{More:2013rma}
  S.~N.~More, A.~Ekstr\"om, R.~J.~Furnstahl, G.~Hagen and T.~Papenbrock,
  Phys.\ Rev.\ C {\bf 87}, no. 4, 044326 (2013).

\bibitem{Wendt:2015}
  K.~A.~Wendt, C.~Forss\'en, T.~Papenbrock and D. S\"a\"af,
  Phys.\ Rev.\ C {\bf 91}, 061301 (2015).

\bibitem{Odell:2016}
  D. Odell, T.~Papenbrock and L. Platter,
    [arXiv:1512.04851 [nucl-th]].

\bibitem{Kim:2014iea}
  Y.~Kim, I.~J.~Shin, P.~Maris, J.~P.~Vary, C.~Forss\'en and J.~Rotureau,
  Int.\ J.\ Mod.\ Phys.\ E {\bf 23}, 1461004 (2014).

\bibitem{Shirokov:2008jv_prelim}
  A.~M.~Shirokov, A.~I.~Mazur, E.~A.~Mazur and J.~P.~Vary,
  Applied Mathematics and Information Sciences {\bf 3}, 245 (2009).

\bibitem{Shirokov:2008jv} 
  A.~M.~Shirokov, A.~I.~Mazur, J.~P.~Vary and E.~A.~Mazur,
  Phys.\ Rev.\ C {\bf 79}, 014610 (2009).

\bibitem{Mazur:2014_NTSE}
  A.~I.~Mazur, A.~M.~Shirokov, J.~P.~Vary, P.~Maris and A.~I.~Mazur,
  Proceedings of the International Workshop on Nuclear Theory in the Supercomputing Era NTSE-2012,
  edited by A.M. Shirokov and A.I. Mazur, Pacific National University Press, Khabarovsk, Russia, pp.146-162 (2013).

\bibitem{Tanihata:2013jwa}
  I.~Tanihata, H.~Savajols and R.~Kanungo,
  Prog.\ Part.\ Nucl.\ Phys.\  {\bf 68}, 215 (2013).

\bibitem{Caurier:2005rb} 
  E.~Caurier and P.~Navr\`atil,
  Phys.\ Rev.\ C {\bf 73}, 021302 (2006).

\bibitem{Caprio:2012rv} 
  M.~A.~Caprio, P.~Maris and J.~P.~Vary,
  Phys.\ Rev.\ C {\bf 86}, 034312 (2012).

\bibitem{AjzenbergSelove:1988ec}
  F.~Ajzenberg-Selove,
  Nucl.\ Phys.\ A {\bf 490}, 1 (1988).

\bibitem{Tilley:2002vg}
  D.~R.~Tilley, C.~M.~Cheves, J.~L.~Godwin, G.~M.~Hale, H.~M.~Hofmann, J.~H.~Kelley, C.~G.~Sheu and H.~R.~Weller,
  Nucl.\ Phys.\ A {\bf 708}, 3 (2002).

\bibitem{Vaintraub:2009mm}
  S.~Vaintraub, N.~Barnea and D.~Gazit,
  Phys.\ Rev.\ C {\bf 79}, 065501 (2009).

\bibitem{Marcucci:2008mg}
  L.~E.~Marcucci, M.~Pervin, S.~C.~Pieper, R.~Schiavilla and R.~B.~Wiringa,
  Phys.\ Rev.\ C {\bf 78}, 065501 (2008).

\bibitem{Pervin:2007sc}
  M.~Pervin, S.~C.~Pieper and R.~B.~Wiringa,
  Phys.\ Rev.\ C {\bf 76}, 064319 (2007).

\bibitem{Johansson:2016}
  H.~T.~Johansson and C.~Forss\'en,
  SIAM J. Sci. Comput. 38, A376 (2016).

\bibitem{bri}
  L.~Brillouin,
  Act. Sci. Ind. {\bf 71}, 159 (1933).

\bibitem{pillet}
  N.~Pillet, J.-F.~Berger and E.~Caurier,
  Phys.\ Rev.\ C {\bf 78}, 024305 (2008);
  N.~Pillet, V.~G.~Zelevinsky, M.~Dupuis, J.-F.~Berger and J.-M.~Daugas,
  Phys.\ Rev.\ C {\bf 85}, 044315 (2012);
  C.~Robin, N.~Pillet, D.~Pe$\tilde{\rm n}$a Arteaga and J.~F.~Berger,
  Phys.\ Rev.\ C {\bf 93}, no. 2, 024302 (2016).

\end{thebibliography}
\end{document}